\documentclass[useAMS,usenatbib]{mn2e}

\usepackage{graphicx}
\usepackage{times}
\usepackage{amssymb}
\usepackage{natbib}

\usepackage{pgf}
\usepackage{bm}

\def\gsim { \lower .75ex \hbox{$\sim$} \llap{\raise .27ex \hbox{$>$}} }
\def\lsim { \lower .75ex \hbox{$\sim$} \llap{\raise .27ex \hbox{$<$}} }

\begin{document}

\title[Radiation pressure \& photoionization feedback]{Stellar feedback by radiation pressure and photoionization}

\author[Sales et al.]{
\parbox[t]{\textwidth}{
Laura V. Sales$^{1}$\thanks{email: lsales@MPA-Garching.MPG.DE},     
Federico Marinacci$^{2,3}$,
Volker Springel$^{2,3}$ and
Margarita Petkova$^{4,5}$
}\vspace*{0.2cm}\\
$^{1}$ Max-Planck Institute for Astrophysics, Karl-Schwarzschild-Strasse 1, 85740 Garching, Germany\\
$^{2}$ Heidelberger Institut f\"ur Theoretische Studien, Schloss-Wolfsbrunnenweg 35, 69118 Heidelberg, Germany\\
$^{3}$ Zentrum f\"ur Astronomie der Universit\"at Heidelberg, Astronomisches Recheninstitut, M\"onchhofstr. 12-14, 69120 Heidelberg, Germany\\
$^{4}$ Department of Physics, Ludwig-Maximilians-Universität, Scheinerstr. 1,
D-81679 M\"unchen, Germany\\
$^{5}$ Excellence Cluster Universe, Boltzmannstr. 2, D-85748 Garching, Germany
}
\maketitle

\begin{abstract} 
  The relative impact of radiation pressure and photoionization
  feedback from young stars on surrounding gas is studied with
  hydrodynamic radiative transfer (RT) simulations. The calculations
  focus on the single-scattering ({\it direct} radiation pressure) and
  optically thick regime, and adopt a moment-based RT-method
  implemented in the moving-mesh code {\sc arepo}.  The source
  luminosity, gas density profile and initial temperature are varied.
  At typical temperatures and densities of molecular clouds, radiation
  pressure drives velocities of order $\sim 20\, \rm km \, s^{-1}$
  over $1 $-$5 \, \rm Myr$; enough to unbind the smaller
  clouds. However, these estimates ignore the effects of
  photoionization that naturally occur concurrently. When radiation
  pressure and photoionization act together, the latter is
  substantially more efficient, inducing velocities comparable to the
  sound speed of the hot ionized medium ($10$-$15 \, \rm km\,s^{-1}$) on
  timescales far shorter than required for accumulating similar
  momentum with radiation pressure.  This mismatch allows
  photoionization to dominate the feedback as the heating and
  expansion of gas lowers the central densities, further diminishing
  the impact of radiation pressure. Our results indicate that a proper
  treatment of the impact of young stars on the interstellar medium
  needs to primarily account for their ionization power whereas direct
  radiation pressure appears to be a secondary effect. This conclusion
  may change if extreme boosts of the radiation pressure by photon
  trapping are assumed.
\end{abstract}


\section{Introduction}
\label{sec:intro}

Massive stars can dramatically change the environment in which they are
born. Although the most profound transformation awaits their final
explosion as supernova, their impact on the surrounding gas starts
much earlier, mediated by their ionizing radiation. Shortly after
young stars start shining, the emitted photons ionize the surrounding
media, burning hot ionized bubbles into the otherwise cold
neutral gas. The production of these HII regions has interesting
consequences not only for the structure and the dynamics of the gas at
scales of clouds and the interstellar medium (ISM), but potentially
also on much larger galactic/halo scales. The overall importance of
these effects depends strongly on the efficiency with which radiation
couples to the gas, the subject studied in  this paper.

Photons carry energy $E=h \nu$ and momentum $P=h \nu /c$, where $h$ is
Planck's constant, $\nu$ the photon frequency and $c$ the speed of
light.  Both quantities are conserved during photon-atom
interactions. The photon's energy can change the ionization state of
an atom and produce heat through the thermalization of the left-over
energy above the ionization threshold (e.g. energy above $E_0=13.6 \,
\rm eV$ for the case of a neutral hydrogen atom initially in the
ground state). The photon's momentum will be transferred to the atom
too, which thus receives a velocity kick in the direction of the
absorbed photon.

Young massive stars emit mostly in the UV region of the spectrum, with
a frequency averaged mean energy $\left<E\right> \sim 20 \, \rm
eV$. This is enough to ionize the neutral gas and also to heat it up
to temperatures $T\sim 10^4 \, {\rm K}$. The hot bubbles produced
around stars -- the HII regions -- are then hotter than the
surrounding medium, which is characterized by temperatures $T\sim 10^2
\,\rm K$ and below. These temperature differences create
pressure imbalances, resulting in a net acceleration of the gas
\citep[see Chapter 20,][for a detailed description]{Shu1992v2}. This
process, here collectively referred to as ``photoionization'', can induce
significant radial velocities in the gas away from the heating
sources. Early analytical calculations and more recent numerical
simulations have confirmed the importance of photoionization for the
dynamics of molecular clouds
\citep{Whitworth1979,McKee1984,McKee1989,Matzner2002,Dale2005,Krumholz2006,
 Dale2012,Walch2012}.

However, the momentum of the photon is also conserved. An atom of mass
$m$ will receive an additional velocity kick $\Delta V = E/(c \; m)$
per absorption of a photon with energy $E$, {\em independent} of the
global effects induced by photoionization.  The extra momentum acts
like a pressure force and is commonly referred to as ``radiation
pressure''. We can make a simple estimate of this effect in the case
of an optically thick shell of gas in the presence of an ionizing
source with luminosity $L$. The optically thick regime guarantees that
all photons are absorbed at the edge of the shell, which will hence
increase its momentum $P$ according to \citep{Murray2005}:
\begin{equation}
\frac{{\rm d}P}{{\rm d}t} = \frac{GM(r)M_g(r)}{r^2} + \frac{L}{c} ,
\label{eq:dpdt}
\end{equation}
\noindent where $M$ and $M_g$ are the total enclosed mass and
gas mass, respectively.  If one ignores the effects of gravity
described by the first term on the right hand side in Eq.~(\ref{eq:dpdt}),
the velocity of the gas can be written as:
\begin{equation}
V_{\rm shell} = \frac{L \; t}{c \; M_{\rm shell}}
\label{eq:vshell}
\end{equation}
\noindent where $M_{\rm shell}$ is the mass of the shell, which will
in general depend on the underlying density distribution. In the case
of a uniform gas density, $\rho_0$, the velocity induced in the gas
will be \citep{Wise2012}:
\begin{equation}
V_{\rm shell} = tA(r_i^4  + 2At^2)^{-3/4} 
\label{eq:lc_wise}
\end{equation}
\noindent where $A=3L/4\rm\pi \rho_0 c$ and $r_i$ is the initial position
of the shell. \citet{Wise2012} argue that the shell will first form when
ionizations balance recombinations at the Str\"omgren radius,
then $r_i=r_s=(3\dot{N}_\gamma/4 \rm\pi \alpha_B (n_H)^2)^{1/3}$, where
$\dot{N}_\gamma$ is the emission rate of photons, $n_{\rm H}$ the hydrogen
number density and $\alpha_{\rm B}$ the case-B recombination coefficient
\citep{Dopita2003}.

Using typical values for the emission of young massive stars, the
arguments above suggest that radiation pressure alone could drive
winds reaching a few hundred $\rm km\,s^{-1}$ \citep{Murray2005,
  Sharma2011,Hopkins2011, Wise2012}. Also, radiation pressure from a
central AGN could be relevant \citep[e.g. ][]{Haehnelt1995} too.  It
is also noteworthy that momentum cannot be radiated away, unlike
energy in the form of heat, a property that might be beneficial in
numerical simulations where lack of numerical resolution can
artificially boost radiative cooling losses.  These factors have
stimulated a significant interest in the galaxy formation community in
radiation pressure as an alternative to thermal supernova feedback
for regulating star formation and, in particular, for driving
galactic winds.

Because of the technical challenges of including radiative transfer
calculations within galaxy-scale simulations, the effects of radiation
pressure have often been included in a ``sub-grid'' fashion that
attempts to approximate the radiation transport effects
\citep[however, see][]{Wise2012, Kim2013}. The accuracy of these
  recipes depends heavily on the particular assumptions and
  approximations made in each model, and thus it is perhaps not
  surprising that the reported results for radiation pressure effects
  vary considerably, from helping to regulate star formation
  \citep{Wise2012, Agertz2013} to driving galactic scale winds that
  remove gas from the galaxies out the halo and beyond
  \citep{Oppenheimer2006, Aumer2013, Ceverino2013}.

In part, these large differences are rooted in varying assumptions
about the effect of radiation trapping by dust. Dust grains can absorb
photons and re-emit them in the infrared, making more photons
available for absorptions. This ``boosts'' the effect of radiation
pressure by a factor $\tau=\kappa_{\rm IR} \Sigma_{\rm gas}$, which
depends on the dust opacity $\kappa_{\rm IR}$ and the gas column
density $\Sigma_{\rm gas}$. There is no clear consensus on realistic
values for $\tau$; under normal galaxy conditions, some authors have
argued in favor of values in the range $0-30$ \citep[][ and references
  therein]{Hopkins2011}. Strictly speaking, we note that this
mechanism can however not be regarded as direct momentum driving any
more since it depends on the absorption of photons with a given energy
$E$ and their re-radiation, having no link with the momentum $E/c$
originally carried by the source photons \citep[][]{Krumholz2009}.  In
this paper, we will not consider radiation trapping but concentrate on
the momentum-driven feedback given by the interaction of the gas with
the momentum content of each photon. This is sometimes referred to as
the ``single-scattering'' regime.

Massive stars not only emit significant amounts of radiation but 
they also deposit energy into their surroundings via stellar winds, 
whose energy accumulated during the lifetime of the star can match that 
of the final supernova explosions \citep{Castor1975}. Recent studies suggest that 
stellar winds can have a sizable impact on the structure of clouds 
\citep{Harper-Clark2009,Rogers2013}, albeit the rate at which they modify
the dynamics might be slow \citep{Dale2008}. Analytical estimates of the 
total energy budget deposited by massive stars suggest that radiation will
dominate over the input from winds \citep{Matzner2002}, with stellar population models 
suggesting a factor $\sim 100$ higher energy deposition in radiation 
than the expected from supernova explosions \citep{Agertz2013}. Determining 
the efficiency with which this energy couples to the surrounding gas is 
therefore fundamental to understand the very nature of stellar 
feedback.

A proper treatment of radiation-induced winds in galaxies requires a
close look at the physical scales of a few $\rm pc$ typical of stellar
HII regions. The problem is challenging not only because of the large
dynamic range of scales involved, but also because it requires an
appropriate treatment of the radiation transport and its coupling to
the dynamics of the gas. Most of the recent progress in this area has
been made by analytical studies, which suggest that radiation pressure
is small compared to the gas pressure by photoionization
\citep[e.g.][]{Mathews1969,Gail1979,Arthur2004}.  However, this might
not be true for more luminous stars embedded in massive molecular
clouds \citep{Krumholz2009}.  Studying the problem with numerical
simulations is attractive because such studies account for spatially
resolved details normally not considered in the analytical treatments
(which usually relate to the behaviour of a ``shell'' or radius of
interest). Also, numerical simulations can naturally follow the
non-linear coupling between the radiation, the dynamics of the gas,
and the gravitational field created by the total mass distribution.
Here we take a step in this direction.

We present a series of controlled numerical simulations with radiative
transfer of idealized set-ups, with a central ionizing source embedded
in an initially neutral gaseous medium composed of hydrogen.  We
compare the effects of photoionization and radiation pressure in
optically-thick gas under two configurations: $i)$ a constant density
media and $ii)$ an isothermal density profile.  The paper is organized
as follows: our code and radiative transfer implementation are
described in Section~\ref{sec:rt}, the effects of radiation on the gas
are examined in Section~\ref{sec:constant} and ~\ref{sec:clouds}. We
discuss and compare our findings with previous work in
Section~\ref{sec:disc} and conclude by highlighting our most important
results. Additional radiative transfer tests and convergence studies
are presented in the Appendices.

\section{Numerical Simulations with Radiative Transfer}
\label{sec:rt}

In our simulations we use the {\sc arepo} code \citep{Springel2010} to
which we have added a moment-based radiative transfer (RT) module
inspired largely by the scheme presented in \citet[][hereafter
  PS09]{Petkova2009}. In the following, we briefly describe the main
characteristics of the code. We refer the reader to the original
papers for further details.

\subsection{Hydrodynamics and gravity}

{\sc arepo} uses an unstructured moving mesh to discretize and solve
the hydrodynamic Euler equations. The mesh is defined as the Voronoi
tessellation of the computational domain resulting from a set of
mesh-generating points which are allowed to move freely and follow the
local flow velocity. The fluxes across cells are computed with a
second-order accurate Godunov scheme with an exact Riemann solver. A
method for on-the-fly refinement and de-refinement of cells can be
invoked to ensure, for example, that fluid cells have masses that do
not differ by more than a factor of $\sim 2$ from a target mass
resolution \citep{Vogelsberger2012}. The mesh is updated every
timestep, and the geometry of cells is regularized mildly by adding
small steering velocities to the mesh generating point where needed.
A reasonably regular Voronoi mesh reduces reconstruction errors and
allows larger timesteps, it is hence beneficial for the performance and
accuracy of the code. These features allow {\sc arepo} to follow
complex flows in a highly adaptive quasi-Lagrangian manner, and in a
fully Galilean-invariant way where bulk velocities do not introduce
additional advection errors. The performance and suitability of the
code to handle standard fluid problems has been analyzed in several
previous works \citep{Springel2010,Sijacki2012, Bauer2012, Munoz2013}.

The gravity solver is based on a {\sc TreePM} scheme where forces are split
into short- and long-range components, offering a uniformly high force
accuracy and full adaptivity. The backbone of the solver is the same
as in the widely used {\small GADGET-2} code
\citep{Springel2005b}. Despite its novel character, {\sc arepo} has
already been used to tackle a range of astrophysical problems inherent
to galaxy formation and first star formation \citep[][ among
  others]{Vogelsberger2013,Nelson2013,Torrey2013,Marinacci2013,Pakmor2013,Bird2013}.

\begin{figure*} 
\begin{center} 
\includegraphics[width=0.85\linewidth]{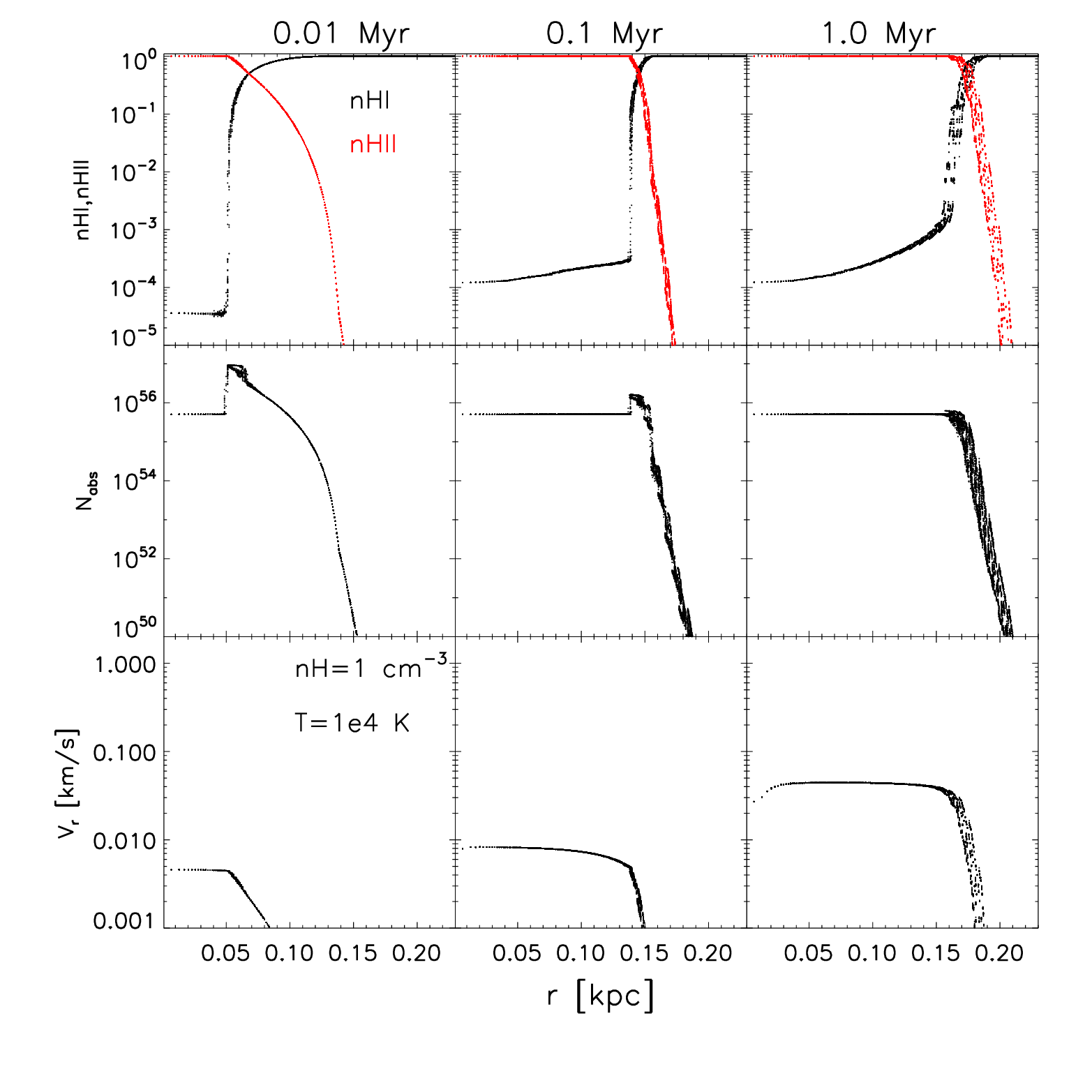} 
\caption{Several quantities of interest in our experiment with a
  constant density box, $n_{\rm H}=1 \, \rm cm^{-3}$, fixed temperature
  $T=10^4 \, \rm K$ and a central source with luminosity $L=10^6 \,
  L_\odot$.  {\it Upper row:} Time evolution of the ionization front in
  a thin slab around the source. Dots correspond to individual cells, and
  black and red indicate the fraction of neutral and ionized gas,
  respectively. {\it Middle row:} Number of photons absorbed in a
  given time-step (arbitrary units). Absorptions occur preferentially
  near the edge of the ionized region at early times but are
  distributed uniformly within the Str\"omgren sphere after
  the recombination time is reached; $t_{\rm rec} \sim 0.125 \, \rm Myr$
  for this set up. {\it Bottom row:} Velocity profiles of the gas due
  to radiation pressure. Momentum is deposited only within the ionized
  sphere.}
\label{fig:otvet_ngamma_sigma}
\end{center}
\end{figure*}

\subsection{The radiative transfer module}

We have implemented in {\sc arepo} a new version of the radiative
transfer module introduced in PS09 and originally developed for {\sc
  gadget}. The structural similarity of both codes allowed a
relatively straightforward adaptation with only a small number of
changes.

Briefly, we solve the moments of the radiative transfer equation using
the Eddington tensor approximation as a closure relation.  The
Eddington tensor is estimated under an optically thin approximation,
following the ideas of the optically-thin variable Eddington tensor
(OTVET) approach of \citet{Gnedin2001}. More specifically, the code
solves the following equation:
\begin{equation}
\frac{\partial J_\nu}{\partial t}=\frac{c}{a^2}\frac{\partial}{\partial
x_j}\left( \frac{1}{\hat{\kappa}_\nu}\frac{\partial J_\nu h^{ij}}{\partial
x_i} \right) - c\hat{\kappa}_\nu J_\nu +  c j_\nu ,
\label{eq:rt1}
\end{equation}
\noindent where $J_\nu$ is the mean intensity, $a$ the expansion
factor (for cosmological runs), $j_\nu$ the emission coefficient and
$\hat{\kappa_\nu}$ the comoving absorption coefficient.  $h^{ij}$ is
the Eddington tensor, which is related to the third moment of the
radiation intensity, the radiation pressure $P^{ij}_\nu$, through the
relation $P^{ij}_\nu = J_\nu h^{ij}$. Eq.~(\ref{eq:rt1}) describes the
evolution of the intensity of the radiation field at a given point due
to photon-conserving radiation transport via anisotropic diffusion
(first term right hand side), and the influence of source and sink
terms that are described by the absorption and emissivity terms on the
right hand side.

Solving this equation requires the knowledge of the Eddington tensor,
which is, {\it a-priori} unknown. Following Gnedin \& Abel (2001), we invoke
an optically-thin approximation to compute the tensor, which in
practice involves an inverse distance square law to all sources,
similar to gravity. This approximation neglects possible absorptions
occurring in between the sources and the cell of interest, but it is
in most situations good enough to define the primary {\it direction}
of photon propagation accurately. For instance, in the case of a
single source (e.g. a star), the corresponding dominant eigenvector of
the Eddington tensor points radially away from the location of the
star, as expected for photons streaming freely from a central source
(see also Fig.~\ref{fig_rt:t1_img}).

Originally, PS09's numerical implementation of the method was designed
to work with an SPH framework, with the discretization scheme and
numerical solvers chosen accordingly (see Sec. 4 in PS09). We have
adapted the techniques to our moving-mesh approach with a deliberately
minimum number of changes in the radiative transfer scheme. In {\sc
  arepo}, the information for each cell can be associated with the
mesh-generating points, allowing them to be interpreted as the
corresponding quantities for a fiducial SPH particle, facilitating the
task. We only need to introduce the concept of a smoothing length $h$
to {\small AREPO} (which is otherwise unnecessary) such that kernel
interpolants can be defined in the same way as done in SPH codes like
{\sc gadget}. Once the smoothing lengths are computed by searching for
the closest 64 neighboring mesh-generating points (which represent the
cells), we follow the exact same discretized equations for radiative
transfer as proposed originally in PS09.

For the time integration of the radiative transfer equation, we adopt
an implicit scheme which is solved with an iterative
conjugate-gradient method, ensuring stability of the diffusion problem
and conservation of photon number.  A flux-limiter $\lambda$ enforces
the condition of sub-luminal transport of photons when the intensity
gradients are large. The flux-limited diffusion version of the scheme
is obtained by adding $\lambda$ as a multiplicative factor within the
parenthesis of the first term on the right-hand side of
Eq.~(\ref{eq:rt1}). Based on carrying out several tests problems, we found
that a flux-limiter of the form proposed by
\citet{Levermore1981} provides accurate results (see Appendix A). In
      {\sc arepo} we then use:
\begin{equation}
\lambda = \frac{2R}{6 + 3R + R^2} ,
\label{eq:fluxlimiter}
\end{equation}
\noindent where $R = |\nabla n_\gamma|/ (\kappa\, n_\gamma)$ and $n_\gamma$
gives the number density of photons. We use the same chemical
network described in PS09 to track the changes in ionization state of the gas.

The code can be used in a mono- or multi-frequency mode; in the
latter, the emission is approximated by a black-body spectrum of a
given temperature $T_{\rm BB}$, and decomposed in four different
frequency bins with boundaries $\nu = [13.6,~24.4,~54.4,~70.0]\,{\rm
  eV}$. In the multi-frequency mode, Eq.~(\ref{eq:rt1}) is solved
independently for each bin, accounting for different absorption
cross-sections, photon densities and opacities per bin. The code
handles a mixture of hydrogen and helium gas, but for simplicity we
use pure hydrogen in our experiments below.

\subsection{Photoionization and radiation pressure}

The temperature of the gas is followed by considering several
mechanisms of cooling and heating. The treatment of cooling includes
processes such as recombination cooling, collisional ionization
cooling, collisional excitation cooling and Bremsstrahlung
cooling. All rates are taken from \citet{Cen1992} and are summarized
in the Appendix A of PS09.

For hydrogen-only gas, irradiated by a source of photons with frequency
$\nu$, the photoheating rate is given by
\begin{equation}
\Gamma = n_{\rm HI} \int {\rm d}\Omega \int_{\nu_o} ^ \infty {\rm d}\nu\, \frac{I_\nu
\sigma_\nu}{h\nu} (h\nu - h\nu_o) ,
\label{eq:photoheating}
\end{equation}
\noindent where $\rm d\Omega$ is the differential solid angle and
$\sigma_\nu$ gives the absorption cross-section at frequency $\nu$. 

Notice that a source emitting photons at exactly the ionization
frequency, $\nu=\nu_0$, will not produce any heating in the
gas. However, photoheating becomes important when harder photons are
considered. For the conditions used in our experiments, heating will
always dominate over cooling.

The temperature changes resulting from ionizing radiation can produce
pressure gradients, inducing net motions of the gas. Note that our
radiative transfer scheme is fully coupled to the hydrodynamical
solver which guarantees a self-consistent update of the gas dynamics
due to photoionization.

By means of the radiative transfer module, we track the number of photon
absorptions per cell, allowing the computation of the momentum deposition
into gas by radiation of energy $E$ as:
\begin{equation}
\Delta V_{\rm cell} = \frac{N_{\rm abs} \, E}{c \; m_{\rm cell}},
\label{eq:velcell}
\end{equation}
\noindent where $\Delta V_{\rm cell}$ is the modulus of the velocity
kick given to a cell with mass $m_{\rm cell}$ that absorbs a net
number of photons $N_{\rm abs}$ in a given time-step. The kick is
directed radially outwards from the source.

The radiative transfer module is robust to changes in time-step and
smoothing lengths. It takes full advantage of the gravity-tree
structure to compute the Eddington tensor, minimizing the introduced
overhead. The strongest advantage of this approach, not really
exploited in this work, is its independence from the number of
sources, making it especially useful for applications where this
number is large, such as cosmic reionization. On the other hand, being
a photon diffusion scheme, shadows are not cast properly, compromising the
accuracy when this is important. To check our implementation, we have
carried out the set of standard radiative transfer tests suggested in
\citet{Iliev2006,Iliev2009}, obtaining satisfactory results. We
summarize the tests in Appendix A.

\begin{table} 
\begin{center}
  \caption{Summary of properties for constant density runs. Columns
    show the density, temperature, side length of the box, Str\"omgren radius and
    recombination time. All numerical experiments were initialized with $64^3$
    cells. The central source emits with luminosity $L=10^6 \, \rm L_\odot$.}
\begin{tabular}[width=0.85\linewidth,clip]{|l|c|rcc}
\hline
 $n_{\rm H}$ & $T$ & $L_{\rm box}$  & $r_s$ & $t_{\rm rec}$ \\ 
$[\rm cm^{-3}]$ & [K] & [pc]  & [pc] & [$10^3\,\rm yr$] \\ 
\hline
0.1 & $10^4$ & 4000 & 808.2 & 1218.3 \\
1.0 & $10^2$ & 200 & 47.4 & 2.43 \\ 
1.0 & $10^4$ & 700 & 174.4 & 121.83 \\
50  & $10^4$ &  50 & 12.88 & 2.44 \\
100 & $10^2$ & 9 & 2.20 & 0.024   \\ 
100 & $10^4$ & 30 & 8.12 & 1.22  \\
\hline
\hline
\end{tabular}
\label{tab:table1}
\end{center}
\end{table}

\section{Radiation pressure in a constant density medium}
\label{sec:constant}

\subsection{Radiation pressure acting alone}

Our first series of experiments investigates the effect of radiation
pressure in gas at constant density $n_{\rm H}$ and
temperature $T$. Table~\ref{tab:table1} summarizes the different
initial conditions, where we have adjusted the length of the box to
keep the resolution fixed in units of the local Str\"omgren sphere. At
$t=0$, the gas is fully neutral, and we switch on a central source of
ionizing photons with luminosity $L=10^6 \, L_\odot \simeq 1.8 \times
10^{50} \, \rm photons/s$. For simplicity, in this Section we will
first assume that all photons are emitted at the hydrogen ionization
energy $E_0=13.6 \; \rm eV$. This maximizes the number of ionizations
at a given luminosity, and therefore, the deposition of momentum into
the gas (see Sec.~\ref{sec:clouds} for a different source spectral
shape). Notice that the assumption of a monochromatic source at $E_0$
means that there are no heating sources due to radiation in these runs and
therefore the internal energy per unit mass of the cells, $u$, 
experiences no change due to the presence of a luminous source.
This translates into an approximately constant temperature in the cells, besides
a reduction by a factor of 2 due to the change in molecular weight of 
ionized hydrogen ($T \propto \mu u$).
Although this is a very special set-up, it highlights in a clean way the
effects of the different variables at play.

\begin{figure} 
\begin{center} 
\includegraphics[width=84mm]{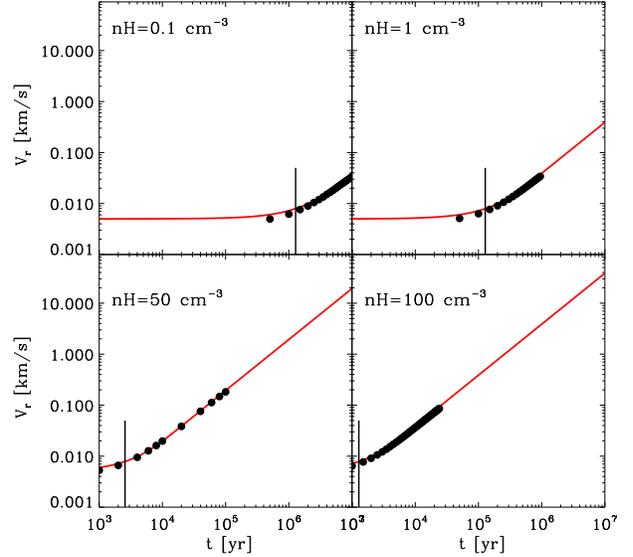} 
\caption{Velocity of the gas measured in simulations with different
  densities, $n_{\rm H}=0.1$, $1$, $50$, and $100 \, \rm cm^{-3}$,
  fixed temperature $T=10^4\, \rm K$ and luminosity $L=10^6\,
  L_\odot$, as a function of time (black dots). $V_r$ corresponds to
  the mass-weighted radial velocity within the instantaneous
  ionization radius $r_{\rm ion}$. The behaviour is well described by
  simple analytical arguments (red lines) where $V_r = L/(c \; m_{\rm
    ion})$ and $m_{\rm ion}$ is the mass within the ionized
  sphere. Short vertical lines show the recombination time $t_{\rm
    rec}$ for each simulation. The gas velocity stays constant as the
  ionized region expands and starts to increase monotonically after
  the Str\"omgren sphere is almost established at $t_{\rm rec}$.}
\label{fig:lc_diffdens}
\end{center}
\end{figure}

Our simulations follow the ionization and recombinations of photons in
the gas via the radiative transfer module described in
Section~\ref{sec:rt}.  Right after the start of the simulation, the
source starts to ionize the surrounding gas, creating an ionized
bubble whose boundary, the ionization radius $r_{\rm ion}$, expands
with time according to the analytic estimate
\begin{equation}
r_{\rm ion} =r_s (1-\exp[-t/t_{\rm rec}])^{1/3}  , 
\label{eq:ri_time}
\end{equation}
\noindent where $r_s$ is the Str\"omgren radius introduced in
Sec.~\ref{sec:intro}, and $t_{\rm rec} = (n_{\rm H} \alpha_{\rm B})^{-1}$ gives
the recombination time.  

In the top row of Fig.~\ref{fig:otvet_ngamma_sigma}, we show the time
evolution of the neutral (black) and ionized (red) fraction as a
function of distance for a fiducial run with density $n_{\rm H} = 1 \,
\rm cm^{-3}$ and temperature $T=10^4 \, \rm K$. Notice that the
assumption of a monochromatic source creates a very sharp transition
between the inner ionized region and the external neutral gas. For the
conditions in this run, the recombination time is $t_{\rm rec}=0.125
\, \rm Myr$ and the size of the Str\"omgren radius is $r_s \sim 175 \,
\rm pc$ (see Table~\ref{tab:table1}), in good agreement with the
ionization profiles in our simulation.

The middle row in Fig.~\ref{fig:otvet_ngamma_sigma} illustrates the
number of absorptions in a time-step around $t=0.01, \, 0.1 \, {\rm
  and} \, 1.2 \, \rm Myr$ (left to right, respectively). At early
times, photons are absorbed preferentially at the edge of the ionized
region, as intuitively expected in the optically thick
regime. However, this ``shell-like'' feature disappears relatively
quickly, and at $t \sim t_{\rm rec}$ the absorptions are
distributed more or less uniformily within the whole ionized
sphere. Notice that this behaviour is quite different from the one
expected in the optically thick-shell scenario discussed in
Sec.~\ref{sec:intro}, where all photons are absorbed at the position
of the shell. Instead, the radiative transfer calculation shows that
photons are being absorbed rather homogeneously within the HII region,
as is necessary to keep the gas ionized.

As a result of these absorptions, the gas builds up a net outward
velocity with a profile shown in the bottom row of
Fig.~\ref{fig:otvet_ngamma_sigma}.  Because momentum cannot be
radiated away, the radial velocity $V_r$ at a given distance
monotonically increases with time as more and more photons get
absorbed.  As expected, the velocity drops to zero in regions not 
reached by any radiation.

\begin{figure} 
\begin{center} 
\includegraphics[width=84mm]{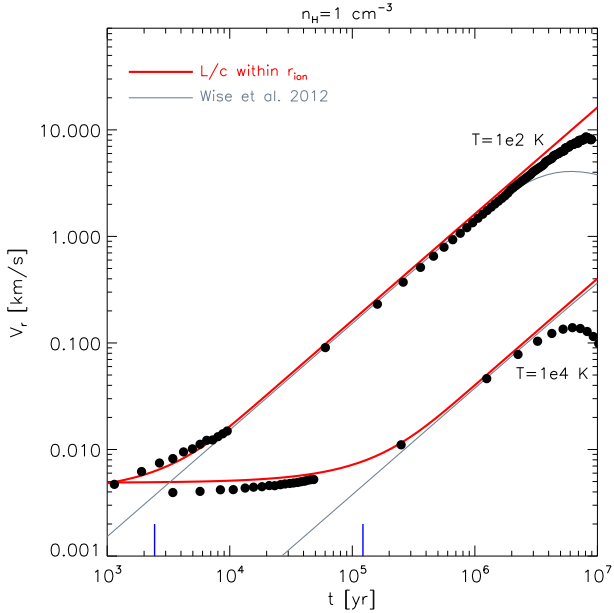} 
\caption{Velocity as a function of time for gas with the same initial
  density $n_{\rm H}=1 \, \rm cm^{-3}$ but different {\it fixed}
  temperatures: $T=10^2 \, \rm K$ (top) and $T=10^4 \, \rm K$
  (bottom).  The velocities acquired via radiation pressure become
  larger the colder the gas is (see text for more details).  Results
  measured in the simulations (black dots) agree well with analytical
  estimations from Eq.~(\ref{eq:lc}) (red curve) at both temperatures.
  Gray lines correspond to the predictions in the case of an optically
  thick shell (Wise et al. 2012), which coincides with our analytic
  estimate for $t > t_{\rm rec}$. At later times, mass entrainment
  will slow down the gas, an effect not considered in
  Eq.~(\ref{eq:lc}) but which can be seen in the thick shell
  formulation as a turn over of the gray curve at $t \sim 4 \times
  10^6\,\rm yr$. The true behaviour of the gas at these late stages
  will lie somewhere between our analytic estimate and the
  optically thick shell formulation.}
\label{fig:lc_temp}
\end{center}
\end{figure}

Figure~\ref{fig:otvet_ngamma_sigma} suggests that the momentum input
by radiation is distributed more or less uniformily within the ionized
region. For a constant density $n_{\rm H}$, we can then compute the expected
velocity of the gas analytically:
\begin{equation}
V_r(t)= L t/(c \; m_{\rm ion}),
\label{eq:lc}
\end{equation}
\noindent where $V_r$ and $m_{\rm ion}$ are the radial velocity and
mass of the gas within the ionized region at time $t$ and $L$ the
luminosity of the source.

Fig.~\ref{fig:lc_diffdens} shows excellent agreement between this
simple estimate and the results of the radiative transfer code for a
variety of gas densities. Interestingly, $V_r$ shows two regimes
established around the recombination time: an early phase where the gas
velocity remains approximately constant (as the ionizing front carves its way
into the neutral gas) and a later regime for $t>t_{\rm rec}$ where the
velocity of the gas increases linearly with time. This behaviour can
be understood in terms of the size of the ionized sphere. When
$r_{\rm ion} < r_s$, the mass in which the momentum is deposited
increases with time while the ionization front runs to its final
equilibrium value $r_{\rm ion}=r_s$. Once the equilibrium between
ionizations and recombinations is reached at $t\sim t_{\rm rec}$,
$m_{\rm ion}$ remains stationary and we expect a linear dependence of $V_r$ with
time, as in Eq.~(\ref{eq:lc}).

The different panels in Fig.~\ref{fig:lc_diffdens} indicate that the
gas velocity becomes larger with increasing gas density. This is because for high
$n_{\rm H}$, the same momentum is being distributed within a
smaller region containing less mass than in a more diffuse media;
$r_{\rm ion}^{3} \propto n_{\rm H}^{-2}$ and $m_{\rm ion} \propto
n_{\rm H}^{-1}$.  Radiation pressure is therefore most relevant -- in
terms of achieving large velocities -- in high density gas.

Temperature also determines the size of the ionized region and can
therefore have an impact on $V_r$.  At fixed density and luminosity,
low temperature gas recombines more efficiently, with a recombination
coefficient $\alpha_{\rm B} (T) \propto (T/10^4)^{-0.85} \rm cm^3\, s^{-1}$
\citep{Dopita2003}. Figure~\ref{fig:lc_temp} shows that by reducing
the temperature from $T=10^4 \, \rm K$ to $10^2 \, \rm K$, the radial
velocity of the gas increased by a factor $\sim 50$, corresponding to
the different $\alpha_{\rm B}$ in each case\footnote{Although
  we show the results for a single gas density $n_{\rm H}=1 \; \rm
  cm^{-3}$, we have explicitly checked that the same trend with
  temperature is observed for other values of $n_{\rm H}$.}.  As before, we
find very good agreement between the results of the numerical
simulations with radiative transfer (black dots) and predictions from
the analytical estimate in Eq.~(\ref{eq:lc}) (thick solid red
lines). The simple assumptions that go into Eq.~(\ref{eq:lc}) seem to
overestimate the gas velocity at early times ($t\ll t_{\rm rec}$), but
the discrepancy is less than $30\%$.

\begin{figure*} 
\begin{center} 
\includegraphics[width=0.475\linewidth]{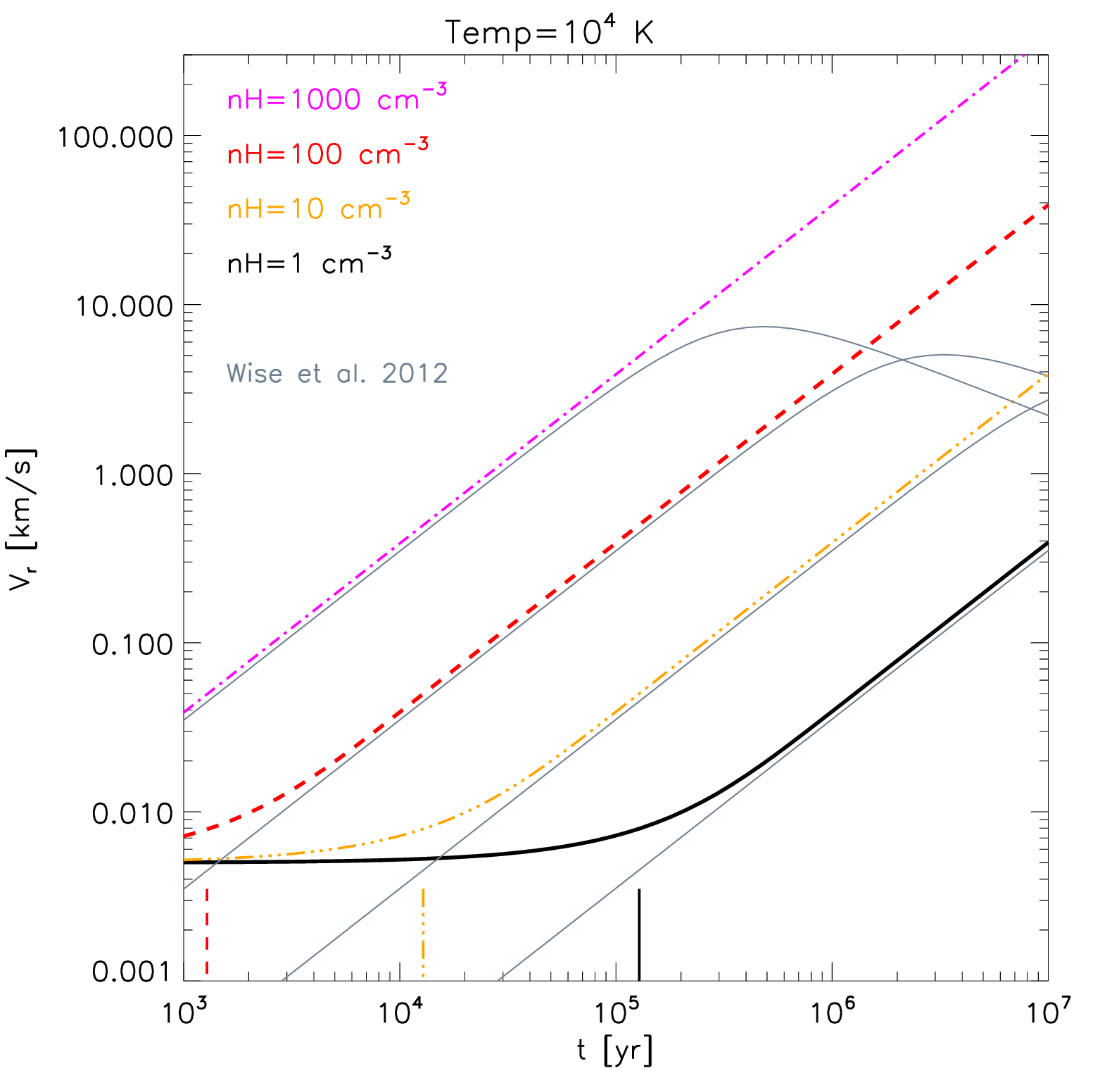} 
\includegraphics[width=0.475\linewidth]{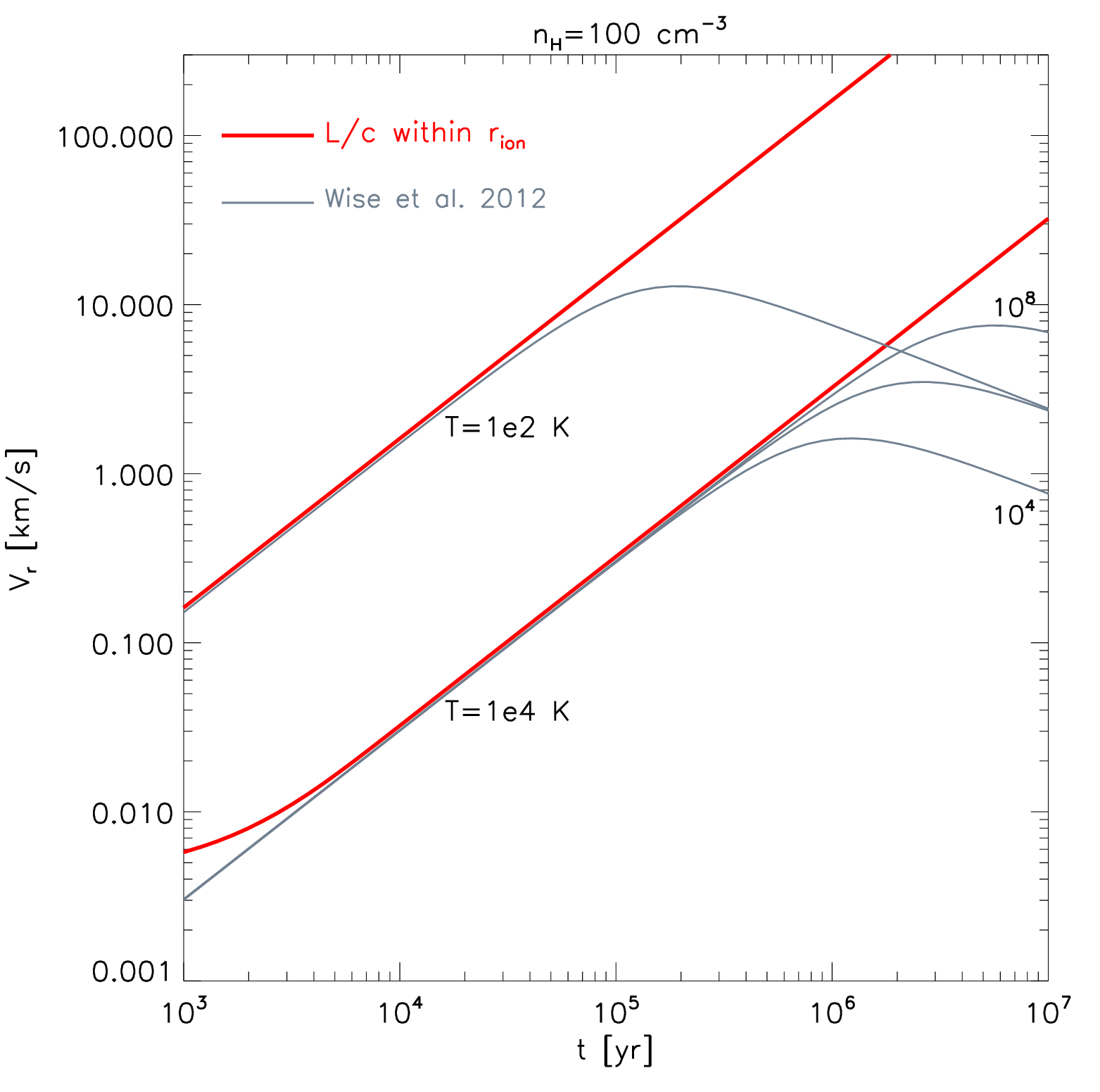} 
\caption{Analytical predictions for gas velocity due to radiation
  pressure. The left panel shows that $V_r$ increases linearly with
  gas density; coloured lines correspond to Eq.~(\ref{eq:lc}) and gray
  ones to the optically thick shell approximation from
  Eq.~(\ref{eq:lc_wise}). The right panel explores, at fixed $n_{\rm
    H}=100 \, \rm cm^{-3}$, the effects of gas temperature and
  luminosity of the source: $V_r$ grows as $\propto (T/10^4)^{-0.85}$
  with decreasing gas temperature but is almost independent of the
  luminosity of the source (lower curves correspond to $L=10^4$,
  $10^6$, and $10^8 \, \rm L_\odot$). This is because more powerful
  sources also have larger Str\"omgren spheres, such that their
  momentum is distributed onto larger masses. These effects cancel out
  in Eq.~(\ref{eq:lc}), producing a single curve irrespective of $L$
  (red curves).  However, the luminosity of the source will determine
  -- towards the end of the lifetime of stars -- when mass entrainment
  takes over in the optically-thick shell formalism, as shown by the
  three gray curves in the $T=10^4 \, \rm K$ set.}
\label{fig:analitic}
\end{center}
\end{figure*}

Despite the fact that the systems behave differently than in the
simplistic optically-thick shell scenario, Fig.~\ref{fig:lc_temp}
shows that the predicted velocities are similar in the two cases. Gray
lines correspond to the analytic estimate of the gas velocity for
an optically-thick shell of gas presented in Eq.~(\ref{eq:lc_wise})
(taken from Eq. 5 in Wise et al. 2012). Although it does not match the
initial phase of expansion of the ionized region, both expressions
agree well in the region of linear growth\footnote{Notice that Wise et
  al. have a typo in their formula for the ionization radius $r_i$
  (Eq. 7) that underestimates their Str\"omgren radius by a factor
  100. The use of this artificially low $r_i$ explains the high
  velocities obtained in their Fig.~1.}. At later times, mass removed
from the inner regions starts to accumulate, slowing down the gas (see
the turn over in the gray curve at $T\sim 5\times10^6 \, \rm yr$ in the $T=100
\, \rm K$ case).  This effect is not taken into account in
Eq.~(\ref{eq:lc}), which continues to increase linearly. The results
of the radiative transfer simulation in the $T=100 \, \rm K$ run
suggest that the true velocity of the gas should fall somewhere in
between both estimates\footnote{The deviation of the $T=10^4 \, \rm K$
  simulation from the analytic curves at $t \sim 4 \times 10^6 \, \rm
  yr$ is only due to our choice of a monochromatic source emitting at
  $E_0$, the specific energy of hydrogen ionization. The lack of heating 
associated to the ionizations combined with a  decrease of density 
in the inner regions creates an inward pressure gradient
that opposes the radiation pressure forces.  However, this is not
expected to be important for real astrophysical sources where ionized
regions are typically hotter than the neutral phase.}.

\begin{figure} 
\begin{center} 
\includegraphics[width=84mm]{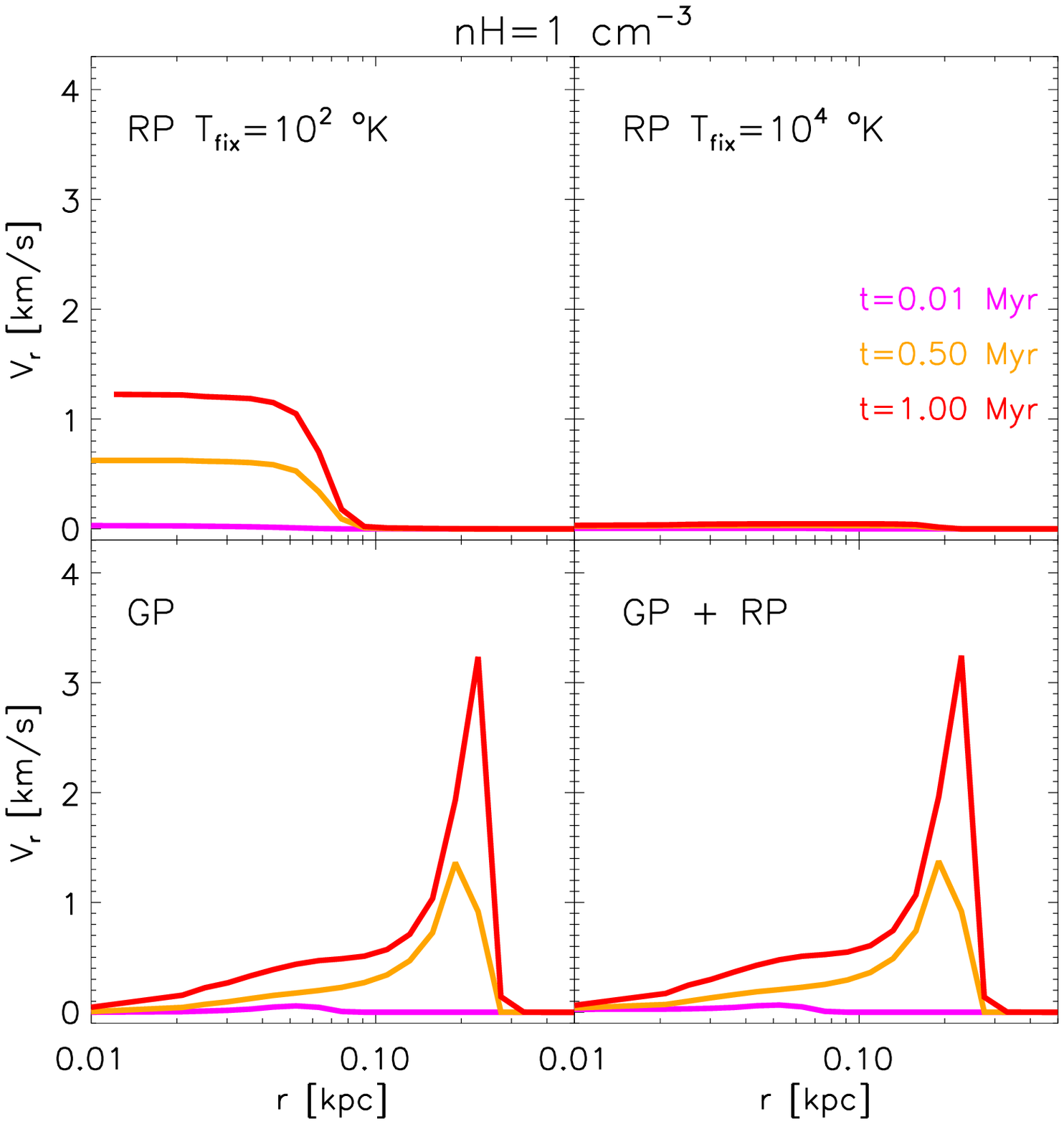} 
\caption{Radial velocity of gas as a function of radius obtained in a
  uniform box with density $n_{\rm H}=1 \, \rm cm^{-3}$ for different
  configurations: radiation pressure only with fixed temperature
  $T_{\rm fix}=10^2 \, \rm K$ (upper left) and $T_{\rm fix}=10^4 \,
  \rm K$ (upper right), photoionization only (bottom left) and
  combined effects of photoionization and radiation pressure (bottom
  right). Simulations in the bottom row where initialized with $T=10^2
  \, \rm K$, and the temperature evolution was followed
  self-consistently to account for heating and cooling
  processes. Colours correspond to three different times as
  labeled. The similarity between the left and right panels in the
  bottom row suggests that the effect of radiation pressure is
  sub-dominant once photoionization is taken into account.}
\label{fig:photion_boxes}
\end{center}
\end{figure}

The good agreement between the radiative transfer calculations and the
analytic estimates from Eq.~(\ref{eq:lc_wise}) and (\ref{eq:lc})
allows us to  gain some intuition about the impact of stellar
radiation on surrounding gas. Figure~\ref{fig:analitic} summarizes the
effect of varying the three primary variables involved in the
radiation pressure problem:
\begin{itemize}
\item {\it Gas density} (left panel): as explained above, $V_r$
  increases linearly with density at fixed temperature and
  luminosity. Notice, however, that for the typical lifetime of
  massive stars, $1-5 \times 10^6 \, \rm yr$, the expected velocities
  due to radiation pressure in gas of temperature $T=10^4 \, \rm K$ are
  small. Even for densities as high as $n_{\rm H}=100 \, \rm cm^{-3}$,
  the characteristic density of molecular clouds, they only reach $V_r
  \sim 5 - 20 \,{\rm km\,s^{-1}}$. This by itself is too low to drive galactic
  winds, but it is large enough to unbind gas clouds in the
  ISM.
\item {\it Gas temperature} (right panel): the gas velocity increases
  with lower temperatures, because the size (and therefore, the mass) of the
  Str\"omgren sphere also decreases for colder gas. Notice, however,
  that the mono-chromatic set-up of our experiments does not yield
  a fully realistic temperature evolution of the ionized region.
\item {\it Luminosity of the source} (right panel): $V_r$ is only
  weekly dependent on the luminosity of the source. Larger
  luminosities translate into larger Str\"omgren spheres and therefore
  a larger $m_{\rm ion}$, in such a way that $V_r$ stays constant. The
  right panel of Fig.~\ref{fig:analitic} shows the velocity of the gas
  at three different luminosities, $10^4$, $10^6$, and $10^8 \, \rm
  L_\odot$, for the $T=10^4 \, \rm K$ case.  The three curves
  perfectly overlap for Eq.~(\ref{eq:lc}) (red line), and only seem to
  differ at late times if we consider the optically-thick shell
  scenario (see three gray curves). This means that for a large
  fraction of the lifetime of the HII region, the total momentum input
  per affected gas mass is independent of luminosity. However, after
  $\sim 1\,{\rm Myr}$ powerful sources might give rise to larger
  velocities, depending on the detailed behaviour of mass entrainment
  and the available time before a supernova explosion.
\end{itemize}

\subsection{Radiation pressure combined with photoionization effects}

For most cases of astrophysical interest, the sources of radiation in the
ISM will be massive stars with spectral emission well approximated by
a black-body spectrum of effective temperature $T_{\rm BB} \sim 10^5
\, \rm K$ (i.e. not monochromatic as assumed above). In this case,
photons will not only ionize but also heat up the gas to a temperature
$T \geq 10^4 \, \rm K$. If the medium was originally at a lower
temperature, this hotter gas within the ionized bubble will create a
pressure gradient outwards, pushing the gas radially away from the
source. In this case, it is the energy as well as the momentum of the
radiation that will have an impact on the dynamics of the gas. The
role played by the radiation pressure should therefore be analyzed in
tandem with the effects of photoionization.

In Fig.~\ref{fig:photion_boxes}, we compare the effects of
photoionization and radiation pressure in detail. We show the gas
velocities as a function of radius in our radiative transfer
simulations where we have considered different mechanisms: radiation
pressure alone at {\it fixed} temperatures $T_{\rm fix}=10^2 \, \rm K$
(top left) and $T_{\rm fix}=10^4 \, \rm K$ (top right),
photoionization alone with an initial temperature $T=10^2 \, \rm K$
(bottom left) and photoionization plus radiation pressure, also with
initial temperature $T=10^2 \, \rm K$ (bottom right). Colored lines
show different times during the evolution, $t=0.01$, $0.5$ and $1 \,
\rm Myr$.  All boxes have an initial constant density $n_{\rm H}=1 \;
\rm cm^{-3}$.  Notice that in the upper row we use a monochromatic
source (i.e. temperature is approximately fixed) whereas the bottom
row uses a black-body spectrum with temperature $T_{\rm BB} = 10^5 \,
\rm K$ and follows the heating and cooling of the gas
self-consistently as described in Sec.~\ref{sec:rt}. We have checked
that the results in the top panels do not change if we use instead a
black-body source and switch-off the heating in the code manually. In
all cases, the luminosity of the source is $L =10^6 \, \rm L_\odot$.

The upper panels in Fig.~\ref{fig:photion_boxes} confirm that
radiation pressure, when acting alone, is more efficient in cold gas.
For example, after a million years, $V_r$ reaches $\sim 1 \, \rm
km\,s^{-1}$ in the $T_{\rm fix}=10^2 \, \rm K$ run compared to
negligible values for $T_{\rm fix}=10^4 \, \rm K$ (see also
Fig.~\ref{fig:lc_temp}). However, tracking the temperature evolution
in a more realistic way shows that photoionization alone can also
yield relevant velocities; reaching in this example $V_r \sim 3 \, \rm
km\,s^{-1}$ in the same time-span (bottom left panel) .

Inspection of the left column of Fig.~\ref{fig:photion_boxes} shows
that photoionization and radiation pressure induce a different
velocity profile; radiation pressure affects only the inner
regions within the Str\"omgren sphere whereas photoionization induces
velocities that peak slightly beyond $r_s$. It is therefore
interesting to look at the joint action of both mechanisms, shown in
the bottom right panel. The similarity between the runs labelled
``GP'' (gas pressure) and ``GP + RP'' (gas pressure + radiation
pressure) in the bottom row suggests that radiation pressure does not
play an important role in the dynamics of the gas for this experiment
once photoionization is  taken into account.

\begin{figure*} 
\begin{center} 
\includegraphics[width=0.475\linewidth]{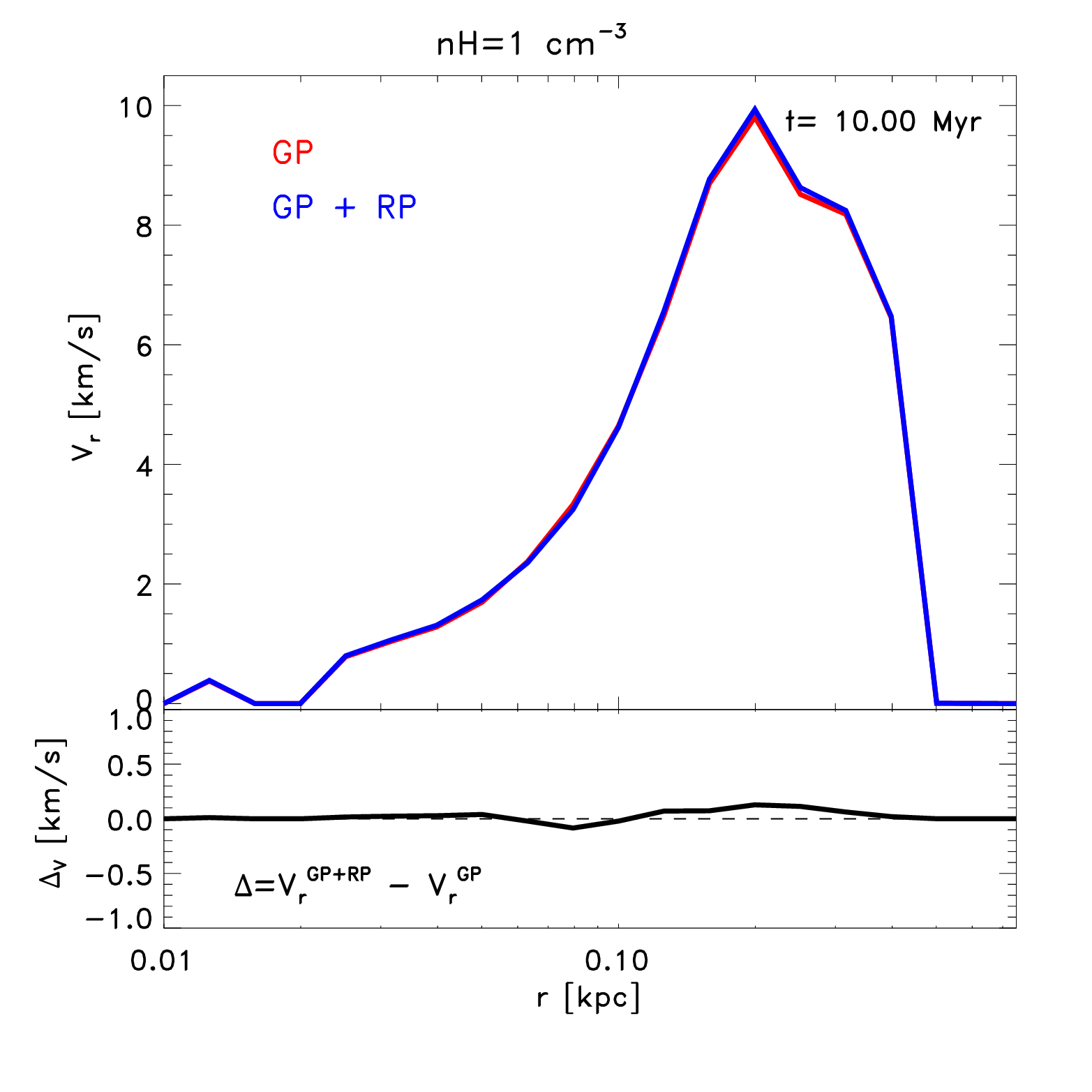} 
\includegraphics[width=0.475\linewidth]{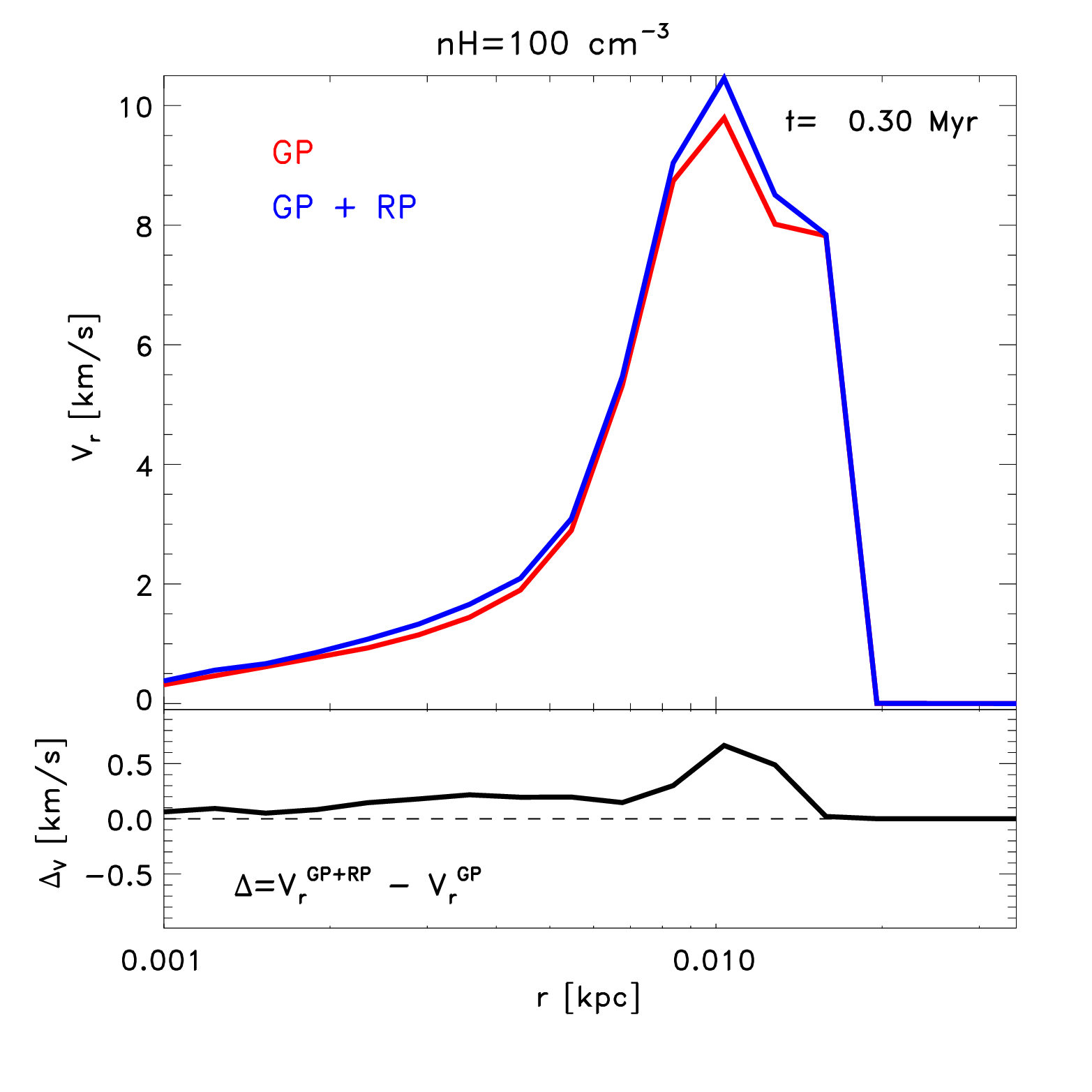} 
\caption{Radial velocity profiles of the gas in the photoionization
  only (red) and photoionization plus radiation pressure (blue) runs
  after $\sim 150 \, t_{\rm rec}$. Both experiments are initialized
  with constant density, $n_{\rm H}=1 \, \rm cm^{-3}$ (left) and
  $n_{\rm H}=100 \, \rm cm^{-3}$ (right), $T=100 \, \rm K$ and the
  temperature is allowed to vary following heating/cooling
  mechanisms. The difference between the red and blue curves is
  strikingly small, supporting a scenario where the dynamics of the
  gas is completely dominated by photoionization (see text for more
  details). The bottom insets show the velocity difference between
  both curves, which corresponds to an increase $\Delta V_r < 0.5\,\rm
  km \,s^{-1}$ for $n_{\rm H}=100 \, \rm cm^{-3}$, and even smaller
  for $n_{\rm H}=1 \, \rm cm^{-3}$.}
\label{fig:photion_veldisp}
\end{center}
\end{figure*}

This is better seen in the left panel of
Fig.~\ref{fig:photion_veldisp}, where we overlap the run with
photoionization alone (red curve) and photoionization plus radiation
pressure (blue) at a much later time $t\sim 10 \, \rm Myr$,
corresponding to approximately $\sim 150\, t_{\rm rec}$, where $t_{\rm
  rec}$ is the recombination time of the ionized gas ($\sim 1.2 \times
10^5 \, \rm yr$). The difference between both curves is negligible, as
shown by the differential curve in the bottom inset. A similar
conclusion is reached for much denser gas, $n_{\rm H}=100 \, \rm
cm^{-3}$ (right panel), where the radiation pressure adds only $\sim
0.5 \, \rm km\, s^{-1}$ in regions of the gas moving at $V_r \sim 9 \,
\rm km \, s^{-1}$ due to photoionization alone. A closer comparison of
the simulations shows that, as soon as ionization starts, it quickly
raises the gas temperature to $T\sim 10^4 \, \rm K$, pushing the
system to the high temperature regime where radiation pressure is less
effective. Therefore, even in high density, initially low temperature
gas, radiation pressure has only a small effect on the dynamics of the
gas. Its influence is mostly overwhelmed by pressure gradients
originating from photoionization.

An important caveat from Fig.~\ref{fig:photion_veldisp} lies in the
time at which we examine the system. For a density $n_{\rm H}=100 \,
\rm cm^{-3}$, $t \sim 250\, t_{\rm rec}$ corresponds to an absolute
time of $\sim 0.3 \times 10^6 \, \rm Myr$, which falls short by a few
million years compared with the expected life time of massive
stars. Unfortunately, we had to stop this simulation because the
shell of material removed from the inner regions reached the border of
the box at $L_{\rm box} = 30 \, \rm pc$ so that we could no longer
track the evolution of the bubble. However, we can confidently place
upper limits on the effect of radiation pressure by using the analytic
estimate from Eq.~(\ref{eq:lc}). The bottom inset in the right panel
of Fig. ~\ref{fig:photion_veldisp} shows that the boost in the gas
velocity due to radiation pressure is $V_r \sim 0.5 \, \rm km\,
s^{-1}$, approximately the velocity predicted for a gas of the same
density and temperature $T=10^4 \, \rm K$ if we consider radiation
pressure only (see Fig.~\ref{fig:analitic}). Using this analytic
calculation, the dashed red curve in Fig.~\ref{fig:analitic} indicates
that radiation pressure will push the gas with speeds no larger than
$\sim 10 \, \rm km \, s^{-1}$ over a period of a few million
years. This is comparable to the velocity caused by photoionization in
a tenth of that time. Moreover, this calculation is an optimistic upper limit,
since the early action of photoionization will lower the density of
the inner regions of the gas, thereby reducing the impact of
radiation pressure.

\begin{figure*} 
\begin{center} 
\includegraphics[width=0.95\linewidth]{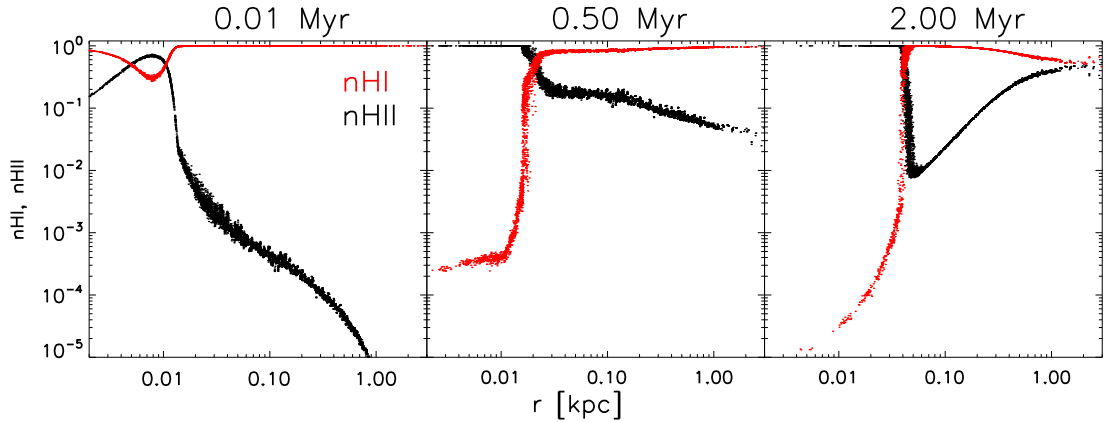} 
\caption{Ionization profiles for the isothermal gas cloud C2 ($\sigma
  = 3 \, \rm km \, s^{-1}$) and a black-body central ionizing source
  with $L=10^6 \, L_\odot$ and $T_{\rm BB}=10^5 \, \rm K$. The
  profiles are more complex than in the constant density case
  (e.g. see Fig.~1). As before, the ionization front moves outwards
  with time and is located at $\sim 4\,\rm pc$ after $2\, \rm Myr$. We
  refer to this case as an ``optically thick'' cloud to distinguish it
  from one in which the ionization front has reached the border of the
  box.}
\label{fig:cloud_otvet}
\end{center}
\end{figure*}

We conclude that for stellar sources surrounded by constant density
gas, photoionization will typically dominate the dynamics of the gas
over direct radiation pressure, unless the density is very high,
$n_{\rm H} \ge 1000 \, \rm cm^{-3}$. Such high densities are not
common on the scale of whole molecular clouds, but could be reached in
small regions close to their cores. We investigate this possibility
below.

\begin{table*} 
\begin{center}
  \caption{Summary of properties of our isothermal-profile gas
    clouds. We list their velocity dispersion $\sigma_{\rm cl}$,
    temperature $T$, central mean density $\left< n_{\rm H} \right>$
    computed within $5 \, \rm pc$, the mass enclosed within $10$ and
    $100 \, \rm pc$, the luminosity of the central source $L$, the
    total end radius of the clouds $R_{\rm tot}$, and the number of
    cells used in the initial conditions $N$. In all the runs, the
    gravitational softening is $\epsilon = 2 \, \rm pc$.}
\begin{tabular}[width=0.85\linewidth,clip]{|l|c|c|c|c|c|c|c|c|}
\hline Label & $\sigma$ & $T$ & $<n_H> (r<5\; \rm pc)$ & $M(r<10\; \rm
pc)$ & $M(r<100\; \rm pc)$ & L & $R_{\rm tot}$ & N\\ & [$\rm km\,
  s^{-1}$] & $[\rm K]$ & $[\rm cm^{-3}]$ & $[ 10^4 \, \rm
M_\odot]$ & $[ 10^4 \, \rm M_\odot]$ & $[\rm L_\odot ]$ & $[\rm kpc]$ &
\\ \hline Cloud 1 (C1) & 1 & 60.57 & 85.4 & 0.18 & 2.27 & $10^4$ & 1 &
$128^3$ \\ Cloud 2 (C2) & 3 & 545.17 & 810.5 & 1.51 & 20.5 & $10^6$ &
3 & $128^3$ \\ Cloud 3 (C3) & 6 & 2180.70 & 3555.6 & 6.5 & 81.0 &
$10^7$& 6 & $256^3$\\ \hline \hline
\end{tabular}
\label{tab:table2}
\end{center}
\end{table*}

\begin{figure*} 
\begin{center} 
\includegraphics[width=0.98\linewidth]{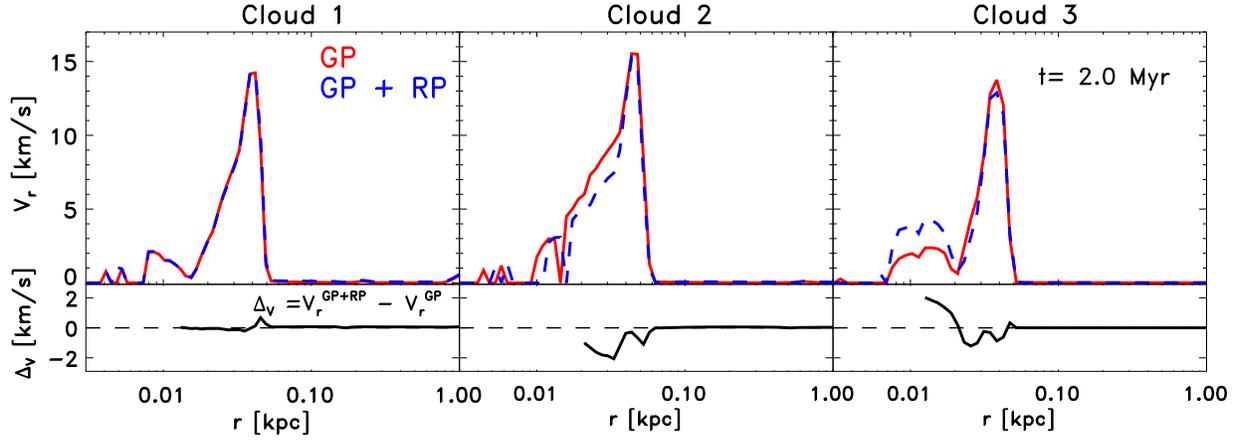} 
\caption{Radial velocity of the gas after $t=2\, \rm Myr$ for
  isothermal-profile clouds with (blue, GP+RP) and without (red, GP)
  radiation pressure. Lower insets show the difference between both
  curves $\Delta_V = V_r^{\rm GP+RP} - V_r^{\rm GP}$, which is very
  small.}
\label{fig:cloud_vel}
\end{center}
\end{figure*}


\begin{figure*} 
\begin{center} 
\includegraphics[width=0.98\linewidth]{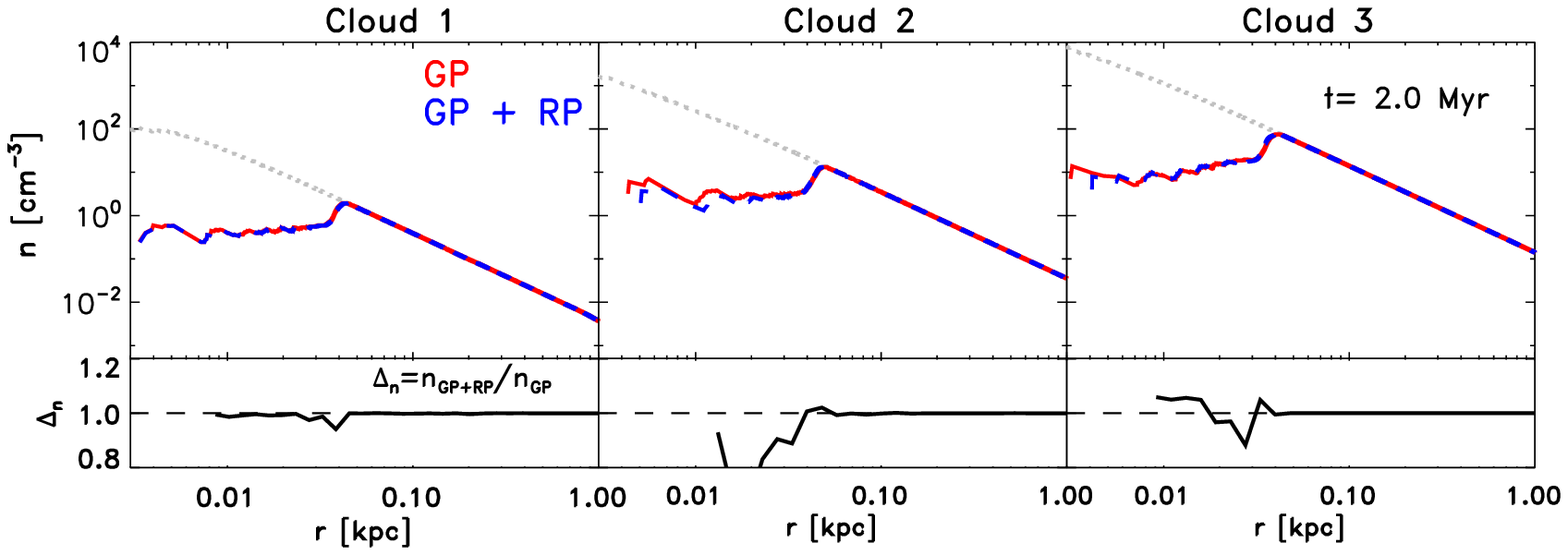} 
\caption{Final density profile for isothermal-profile gas clouds run
  with (blue, GP+RP) and without (red, GP) radiation pressure. The
  results correspond to $t=2 \, \rm Myr$.  The impact of radiation
  pressure is smaller than $\sim 20\%$, as shown by the ratio between
  both profiles $\Delta_n = n_{\rm H}^{\rm GP+RP}/n_{\rm H}^{\rm GP}$
  in the bottom inset panels. }
\label{fig:cloud_rho}
\end{center}
\end{figure*}

\section{Radiation pressure in gas clouds with isothermal profiles}
\label{sec:clouds}

In this Section we relax our assumption of constant density gas and
explore the effects of radiation pressure and photoionization in
isothermal density profiles, $\rho \propto r^{-2}$. This provides a
better description of the gaseous clouds where stars are born
\citep[although the exact structure of molecular clouds can show large
  variations, e.g.][]{Heyer2009}.  A detailed model of radiation
pressure in realistic molecular clouds is out of the scope of this
paper (it would require to include sources of turbulence, resolve
individual star formation, model the effects of metallicity,
etc.). Instead, we aim to highlight, using a series of idealized
experiments, the role played by the density distribution of the gas on
the relative interplay between radiation and photoionization pressure.

\begin{figure*} 
\begin{center} 
\includegraphics[width=0.475\linewidth]{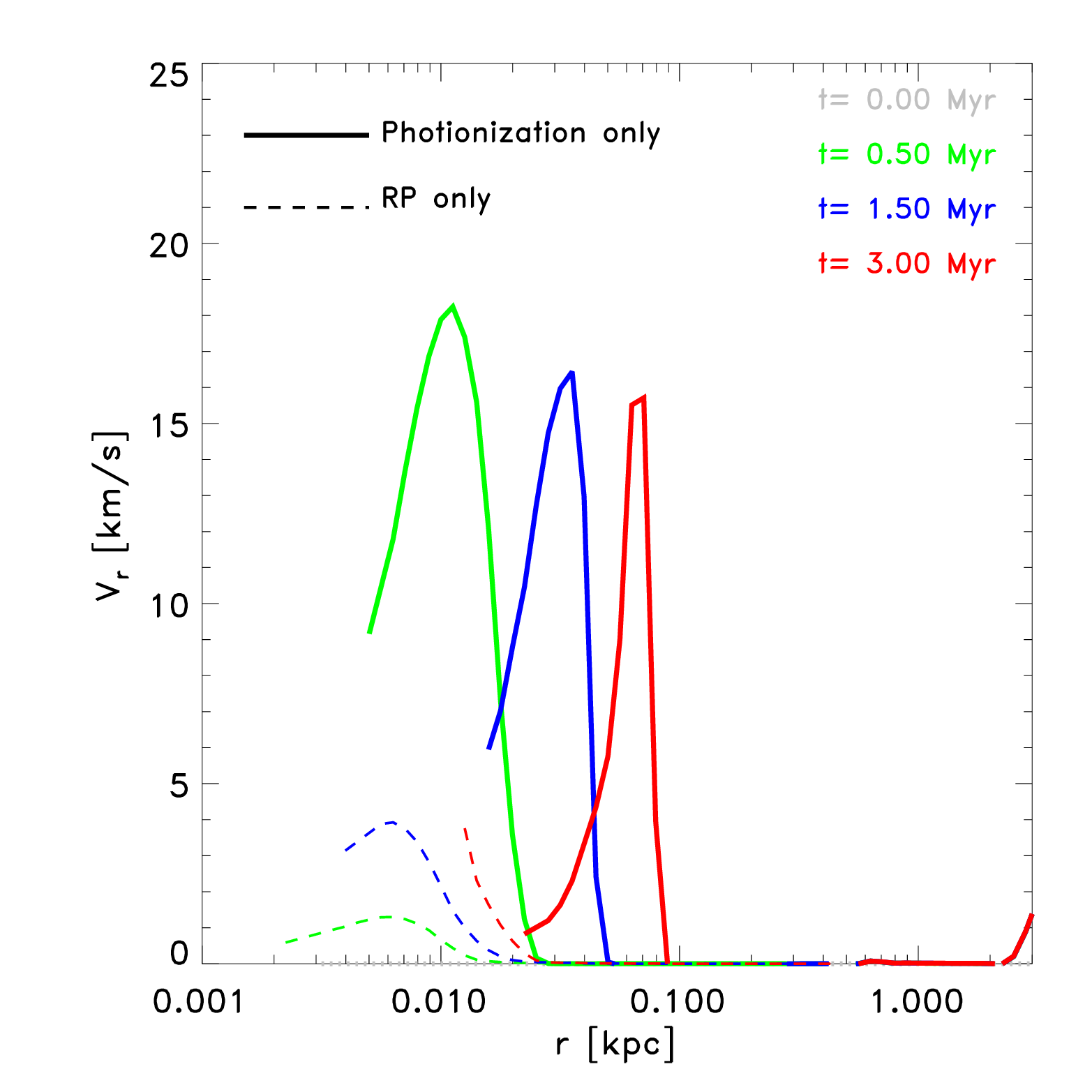} 
\includegraphics[width=0.475\linewidth]{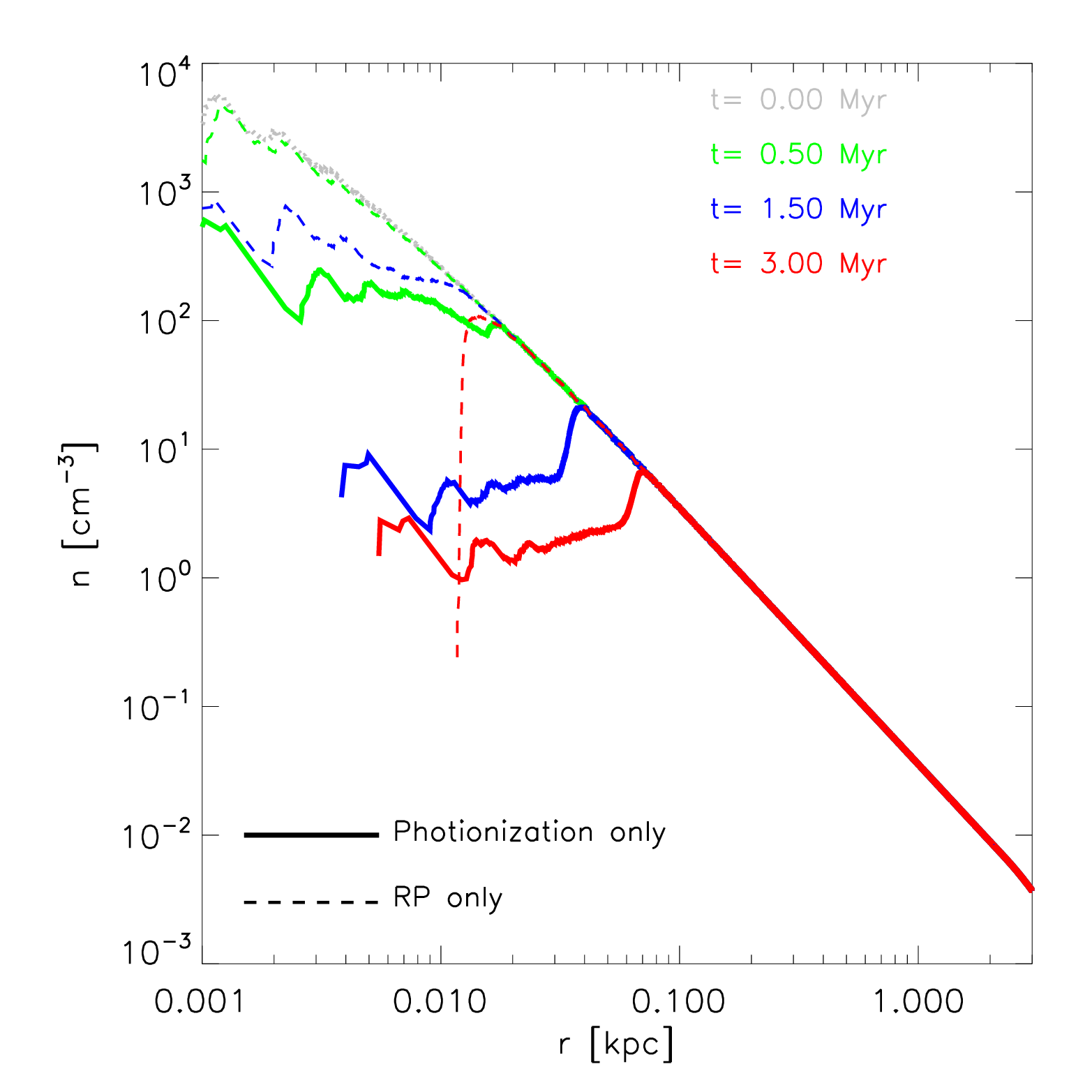} 
\caption{Further diagnostics of the isolated effects of radiation
  pressure compared to photoionization. We show the velocity (left)
  and the density profile (right) of the C2 cloud at different times
  (coloured lines) for runs where only photoionization (solid) or only
  radiation pressure (dashed) are taken into account.  Radiation
  pressure has the ability to modify the velocity and mass
  distribution of the cloud, emptying the inner $ \sim 10 \, \rm pc$
  of the cloud in $\sim 3 \, \rm Myr$. But the time needed for this is
  longer than the time required by photoionization to start affecting
  the medium. The left panel shows that at any given time (same
  colours), photoionization has a larger velocity than radiation
  pressure. Thus, when both effects are considered, photoionization
  sets in quicker, heating and pushing the gas with the result of
  lowering the central densities on a shorter timescale. This in turn
  diminishes the impact of the radiation pressure even further. This
  explains the small differences between red and blue curves in
  Fig.~\ref{fig:cloud_vel} and \ref{fig:cloud_rho}.
\label{fig:clouds_indiv}}
\end{center}
\end{figure*}

We consider three gas clouds C1-C3 with the properties listed in
Table~\ref{tab:table2}. Each cloud was set up in hydrostatic
equilibrium, which defines a relation between total mass, radius,
velocity dispersion and temperature by specifying only one of the four
variables \citep{BinneyTremaine2008}.  Table~\ref{tab:table2} includes
the mass contained within $10$, and $100 \, \rm pc$ (columns 5 and 6)
which may help to relate them to known objects in the local
Universe. Object C1 is an example of a fluffy low mass cloud hosting a
late B-type star, a good representative of the $M \sim 10^3\, \rm
M_\odot$ objects populating the low mass end of the compilation of
molecular clouds by \citet{Heyer2009}.  C2 and C3 are more massive
examples hosting a source with a luminosity of a couple of OB
stars. They are consistent with clouds of mass $\sim 5\times 10^4$,
$5\times 10^5 \, \rm M_\odot$ and sizes of $20$, $50 \, \rm pc$
respectively (Heyer et al. 2008, see also Fig. 3 in Dale et al. 2012
for a graphical display of the data).

The initial conditions of each cloud were evolved in isolation for $1 \,
\rm Gyr$ to verify their dynamic stability. No structural change was
observed. We added a massless central source of ionizing radiation
with a given luminosity (see Table~\ref{tab:table2}) and a black-body
spectrum with temperature $T_{\rm BB}=10^5 \,  \rm K$. The luminosities
were chosen such that all clouds will remain optically thick for a
few million years. We performed radiative transfer simulations with
our code, following the heating and cooling of the gas, the
hydrodynamics and the gravity forces.

The ionization profiles in our clouds are now more complex (see
Fig.~\ref{fig:cloud_otvet}) compared to the simple case shown in
Fig.~\ref{fig:otvet_ngamma_sigma}. This is not only the result of the
declining density profile, but also of the spectral shape of the
ionizing source. The high energy photons from the $10^5 \, \rm K$
black-body spectrum can penetrate deeper into the cloud and ionize
distant regions before the bulk of the $E_0 \sim 13.6 \, \rm eV$
photons fully ionize the interior of the cloud.  It is still possible
to define an ionization radius as the smallest radius where the
neutral fraction is larger than $0.5$.  As mentioned before, we only
focus on clouds where $r_{\rm ion}$ is smaller than the final radius
of the cloud -- the optically thick regime -- and defer an analysis of
fully ionized clouds to future work.

Fig.~\ref{fig:cloud_vel} shows a comparison between the velocity as a
function of distance in our clouds considering photoionization alone
(red) and photoionization plus radiation pressure (blue). We show the
results after 2 Myr of evolution, the time at which the input from
radiation pressure is maximum \citep{Krumholz2009}. As in the previous
section, we again find that adding radiation pressure to our
calculations does not change the gas velocity by a significant
amount. Fig.~\ref{fig:cloud_vel} shows that the velocity in the run
were both radiation pressure and gas pressure effects are considered
is either equal or even smaller than in a run with thermal pressure
alone.  The lower velocities are caused by an enhanced mass removal
from the center of the clouds when radiation pressure is included, but
this effect is small.  Fig.~\ref{fig:cloud_rho} shows that the gas
density in the centers can be lower by up to $\sim 20\%$ in the run
with radiation pressure compared to considering only photoionization.
This is also true for C3, our cloud with the highest mass and most
powerful source, i.e., our most optimistic candidate for radiation
pressure effects. C1, on the other hand, shows no difference between
both runs.  In all cases, photoionization alone generates velocities
$V_r \sim 15 \, \rm km\,s^{-1}$, flattening the inner density profile
of the clouds and dominating the overall evolution in time.

The relatively small impact of radiation pressure in
Fig.~\ref{fig:cloud_vel} and \ref{fig:cloud_rho} is perhaps
surprising: analytic arguments and the results of the radiative
transfer calculation from the previous Section have shown that
radiation pressure can potentially drive significant velocities,
especially for high densities. Why then do our simulated clouds show
such a small effect due to radiation pressure? To answer this question
one has to consider the characteristic timings associated with each
process, photoionization and radiation pressure.

To investigate this in more detail, we run the C2 cloud including {\it
  only} photoionization or {\it only} radiation pressure (no
temperature evolution), leaving everything else unchanged. The results
of this exercise are shown in Fig.~\ref{fig:clouds_indiv}.
Photoionization (solid curves) proceeds relatively quickly, raising
the velocity of the gas above $V_r \sim 15 \, \rm km \, s^{-1}$ in
less than a million years. The peak velocity moves outwards with time,
decreasing slightly in magnitude.  In contrast, radiation pressure
(dashed curves) requires longer times to deposit the momentum in the
media, exciting a maximum velocity of only $V_r \sim 5 \, \rm km \,
s^{-1}$ after 2 or $3\,\rm Myr$.

A similar effect is seen in the density profiles (right panel of
Fig.~\ref{fig:clouds_indiv}). Radiation pressure (dashed curves)
eventually pushes all the gas within $r=10 \, \rm pc$, emptying the
core of the cloud. But it does it slowly, with only minor
modifications in the profile for the first $\sim 1.5 \, \rm Myr$ of
evolution. Instead, photoionization starts emptying the inner regions
considerably faster, with less than $5 \times 10^5 \, \rm yr$ required
to produce appreciable changes in the inner regions of the cloud (see
solid green line). After $3 \, \rm Myr$, photoionization alone has
lowered the density of the core by more than three orders of
magnitudes.

This time delay in the build up of the effects of radiation pressure
with respect to photoionization makes it less relevant than
expected. Radiation pressure needs a comparatively long time period to
deposit the momentum in the gas. The photoionization timescales are
shorter, yielding an early removal of gas from the cloud centre. This
limits the effects of radiation pressure even further, which require
large densities to maximize the momentum input. Radiation pressure in
the single scattering limit is therefore sub-dominant compared with
photoionization effects in the gas for both  mass distributions
explored here, a constant density medium and a declining $r^{-2}$
density profiles.

\section{Discussion and conclusions}
\label{sec:disc}

It is interesting to contrast the results of our radiative transfer
calculations with some analytic predictions. \citet{Krumholz2009} have
shown that the pressures due to radiation and due to photoionization
have different radial dependences, meaning that there is a
characteristic radius $r_{\rm ch}$ within which radiation pressure is
expected to dominate while further out gas pressure takes over. This
characteristic radius is typically small for the average molecular
clouds in the Milky Way, where photoionization might be the dominant
mechanism for cloud destruction (see Walch et al. 2012). However, for
the most dense clouds hosting luminous sources, $r_{\rm ch}$ can reach
a few hundred parsecs, so that radiation pressure might be able to
completely unbind them.

We have used equations 6 and 8 from Krumholz et al. to estimate
$r_{\rm ch}$ for the clouds studied in Sec.~\ref{sec:clouds}. For C1
and C2, we obtain values smaller than our resolved region ($r_{\rm
  ch}=1 \times 10^{-2}$ and $1.5 \, \rm pc$, respectively), thus the
dominance of gas pressure found in our simulations for $r > 5\, \rm
pc$ would be expected in this scenario.  But our most massive cloud,
C3, has $r_{\rm ch}=20 \, \rm pc$, a region well resolved in our
analysis. Nevertheless, for C3 we also find a negligible impact of
radiation pressure on the central density of the
cloud after $t=2 \, \rm Myr$.  This is not
necessarily a disagreement, as the analytical estimates by Krumholz et
al. explore the effect of radiation pressure on the position and
velocity of expansion of the ionization front, and not on the gas in the
clouds, as explored here. The I-front velocity can differ
substantially from the actual bulk velocity induced in the gas (for
example, compare Fig. 12 and 18 in Iliev et al. 2009).  In our
results, radiation pressure does not dominate the gas dynamics and
hence the structure of the cloud at any radius.

In general, we find that radiation pressure is a relatively slow and
inefficient way of coupling radiation to the surrounding gas.  This
can be understood with a simple example. Let us consider the case of
a hydrogen atom that absorbs a single 13.6 eV photon. The momentum
deposited in that atom will push it with velocity $V_r \sim 5.6 \, \rm
m\, s^{-1}$, requiring more than thousand ionizations from the same
direction for a single atom to achieve a modest velocity of $V_r \sim
10 \, \rm km \, s^{-1}$. The core of the problem is that, once the
atom is ionized, it takes some time to recombine and be available to
receive the next velocity kick. That is why radiation pressure is slow
compared to photoionization, which only requires a short time for the
thermalization of the energy above $E=13.6 \, \rm eV$ of a single
ionization event. The inefficiency of direct radiation pressure has
been suggested before
\citep{Mathews1969,Spitzer1978,Arthur2004,Krumholz2009,Kim2013}, but
has not been confirmed rigorously in simple configurations with
appropriate radiative transfer calculations.

Encouragingly, our findings are in good agreement with recent
observational results of HII regions, where gas pressure seems to
dominate over radiation \citep{Lopez2013}, as well as with numerical
modelling of clouds \citep{Dale2013}.  The results of our experiments
have direct consequences for the sub-grid modelling of radiation
pressure at the scale of galaxy simulations. Arguably the most
important lesson is that the ``mass loading'' (i.e.~the mass in which
the photon momentum is deposited) is not a free parameter but is
determined by the size of the ionized region. This agrees with the
recent work by \citet{Renaud2013}, but disagrees with many previous
models where the mass loading was chosen in an {\it ad-hoc} fashion
\citep{Oppenheimer2006, Agertz2013,Aumer2013, Ceverino2013}. This can
have a sizable impact on the effectiveness associated with radiation
pressure feedback in these numerical codes, since by choosing a
sufficiently small mass loading, the velocity given to the gas can be
increased almost arbitrarily (within the constraints of the total
momentum input from the ionizing source).

Under general conditions in the ISM, photoionization due to massive
stars has an impact on the dynamics of the gas that is comparable to
and typically more important than that of direct radiation pressure.
From the point of view of numerical modelling, the ``early feedback''
implementation by \citet{Stinson2013} seems to be a step in the right
direction, in the sense that it actually uses thermal pressure rather
than radiation pressure. A problematic point is however that there is
still lots of freedom when choosing the total energy budget released
by the young stars and how this is distributed from the stars to the
neighboring SPH particles. The effects of photoionization are also
starting to be included in semi-analytical models with promising results
\citep[e.g. ][]{Lagos2013}.

Fig.~\ref{fig:analitic} shows that radiation pressure might be able to
generate competitive velocities $V_r \geq 50 \, \rm km \, s^{-1}$ only
for very high density gas, $n_{\rm H} \geq 1000 \, \rm cm^{-3}$,
although this is uncertain up to the exact slow-down effect due to the
mass accumulation (a shell-like approximation predicts that such
velocities will never be reached). But even neglecting the mass
entrainment, such high densities -- necessary to get very short
recombination times -- are uncommon, occurring only in the most massive
molecular clouds in the Milky Way, or are restricted to particular
environments such as vigorous starbursting galaxies or gas-rich
proto-galaxies in the early Universe. Radiation-pressure driven winds are hence
unlikely to be important for the general locus of galaxies unless we
assume large boosting factors due to radiation trapping in dust
grains.

An assessment of the potential effects of dust on our study is not
straightforward as the presence of dust makes the effective
Str\"omgren radius smaller than the corresponding dust-free case (see
chapter 5 Spitzer 1978). However, we can place some upper limits based
on the predictions from Fig.~\ref{fig:analitic}, which tend to be
conservative as they ignore the effects of photoionization taking
place earlier than the radiation pressure.  At the average density of
molecular clouds, $n_{\rm H}=100 \, \rm cm^{-3}$, we need to boost the
velocities by a factor $\tau \sim 30$ to achieve $V_r \sim 400-600 \,
\rm km \,s^{-1}$ comparable to the escape velocity in $L_*$
galaxies. This would lie at the upper end of conceivable boost factors
according to current estimates \citep{Hopkins2011}, but there have
been some recent discussions about the plausible upper end
\citep{Krumholz2012}.

We recall that we have simulated radiation pressure under highly
idealized conditions with the aim to explore in detail its interplay
with photoionization and their combined dynamical imprint on the
gas. Several caveats are unavoidable when taking such an approach. In
particular, our results deal with spherically symmetric systems and
consider only hydrogen ionization. We also neglect any effect due to
the complicated small-scale structure of the ISM, which we approximate
as a well mixed, isothermal medium. Obvious gaps in our analysis
concern the presence of ``champaign flows'' \citep[e.g. ][]{Yorke1983}
or the acceleration of confined cold clouds within a hot medium
created by ionization. But in turn, the advantage in using idealized
models lies in providing a clean understanding of the interaction
between radiation and the dynamics of the gas under conditions that
are typical in galaxies. This helps to narrow down the regime in which
an inclusion of the effects of radiation pressure is required. The
results presented here might also be useful for developing more
realistic sub-grid models for simulations of galaxy formation.

\section*{Acknowledgements}
\label{acknowledgements}

We are grateful to the anonymous referee for a constructive report that
helped to improve our manuscript.
We would like to thank Rajat Thomas, Martin Haehnelt, Norm Murray, Steffanie Walch,
Thorsten Naab and Simon White for useful and inspiring discussions.
F.M.~acknowledges support by the DFG Research Centre SFB-881 `The
Milky Way System' through project A1.  This work has also been
supported by the European Research Council under ERC-StG grant
EXAGAL-308037 and by the Klaus Tschira Foundation.

\bibliography{master}

\begin{thebibliography}{}

\bibitem[\protect\citeauthoryear{{Agertz}, {Kravtsov}, {Leitner} \&
  {Gnedin}}{{Agertz} et~al.}{2013}]{Agertz2013}
{Agertz} O.,  {Kravtsov} A.~V.,  {Leitner} S.~N.,    {Gnedin} N.~Y.,  2013,
  \apj, 770, 25

\bibitem[\protect\citeauthoryear{{Arthur}, {Kurtz}, {Franco} \&
  {Albarr{\'a}n}}{{Arthur} et~al.}{2004}]{Arthur2004}
{Arthur} S.~J.,  {Kurtz} S.~E.,  {Franco} J.,    {Albarr{\'a}n} M.~Y.,  2004,
  \apj, 608, 282

\bibitem[\protect\citeauthoryear{{Aubert} \& {Teyssier}}{{Aubert} \&
  {Teyssier}}{2008}]{Aubert2008}
{Aubert} D.,  {Teyssier} R.,  2008, \mnras, 387, 295

\bibitem[\protect\citeauthoryear{{Aumer}, {White}, {Naab} \&
  {Scannapieco}}{{Aumer} et~al.}{2013}]{Aumer2013}
{Aumer} M.,  {White} S.~D.~M.,  {Naab} T.,    {Scannapieco} C.,  2013, \mnras,
  434, 3142

\bibitem[\protect\citeauthoryear{{Bauer} \& {Springel}}{{Bauer} \&
  {Springel}}{2012}]{Bauer2012}
{Bauer} A.,  {Springel} V.,  2012, \mnras, 423, 2558

\bibitem[\protect\citeauthoryear{{Binney} \& {Tremaine}}{{Binney} \&
  {Tremaine}}{2008}]{BinneyTremaine2008}
{Binney} J.,  {Tremaine} S.,  2008, {Galactic Dynamics: Second Edition}.
Princeton University Press

\bibitem[\protect\citeauthoryear{{Bird}, {Vogelsberger}, {Sijacki},
  {Zaldarriaga}, {Springel} \& {Hernquist}}{{Bird} et~al.}{2013}]{Bird2013}
{Bird} S.,  {Vogelsberger} M.,  {Sijacki} D.,  {Zaldarriaga} M.,  {Springel}
  V.,    {Hernquist} L.,  2013, \mnras, 429, 3341

\bibitem[\protect\citeauthoryear{{Castor}, {McCray} \& {Weaver}}{{Castor}
  et~al.}{1975}]{Castor1975}
{Castor} J.,  {McCray} R.,    {Weaver} R.,  1975, \apjl, 200, L107

\bibitem[\protect\citeauthoryear{{Cen}}{{Cen}}{1992}]{Cen1992}
{Cen} R.,  1992, \apjs, 78, 341

\bibitem[\protect\citeauthoryear{{Ceverino}, {Klypin}, {Klimek},
  {Trujillo-Gomez}, {Churchill}, {Primack} \& {Dekel}}{{Ceverino}
  et~al.}{2013}]{Ceverino2013}
{Ceverino} D.,  {Klypin} A.,  {Klimek} E.,  {Trujillo-Gomez} S.,  {Churchill}
  C.~W.,  {Primack} J.,    {Dekel} A.,  2013, ArXiv e-prints 1307.0943

\bibitem[\protect\citeauthoryear{{Dale} \& {Bonnell}}{{Dale} \&
  {Bonnell}}{2008}]{Dale2008}
{Dale} J.~E.,  {Bonnell} I.~A.,  2008, \mnras, 391, 2

\bibitem[\protect\citeauthoryear{{Dale}, {Bonnell}, {Clarke} \& {Bate}}{{Dale}
  et~al.}{2005}]{Dale2005}
{Dale} J.~E.,  {Bonnell} I.~A.,  {Clarke} C.~J.,    {Bate} M.~R.,  2005,
  \mnras, 358, 291

\bibitem[\protect\citeauthoryear{{Dale}, {Ercolano} \& {Bonnell}}{{Dale}
  et~al.}{2012}]{Dale2012}
{Dale} J.~E.,  {Ercolano} B.,    {Bonnell} I.~A.,  2012, \mnras, 424, 377

\bibitem[\protect\citeauthoryear{{Dale}, {Ngoumou}, {Ercolano} \&
  {Bonnell}}{{Dale} et~al.}{2013}]{Dale2013}
{Dale} J.~E.,  {Ngoumou} J.,  {Ercolano} B.,    {Bonnell} I.~A.,  2013, \mnras,
  436, 3430

\bibitem[\protect\citeauthoryear{{Dopita} \& {Sutherland}}{{Dopita} \&
  {Sutherland}}{2003}]{Dopita2003}
{Dopita} M.~A.,  {Sutherland} R.~S.,  2003, {Astrophysics of the diffuse
  universe}

\bibitem[\protect\citeauthoryear{{Gail} \& {Sedlmayr}}{{Gail} \&
  {Sedlmayr}}{1979}]{Gail1979}
{Gail} H.~P.,  {Sedlmayr} E.,  1979, \aap, 77, 165

\bibitem[\protect\citeauthoryear{{Gnedin} \& {Abel}}{{Gnedin} \&
  {Abel}}{2001}]{Gnedin2001}
{Gnedin} N.~Y.,  {Abel} T.,  2001, \na, 6, 437

\bibitem[\protect\citeauthoryear{{Haehnelt}}{{Haehnelt}}{1995}]{Haehnelt1995}
{Haehnelt} M.~G.,  1995, \mnras, 273, 249

\bibitem[\protect\citeauthoryear{{Harper-Clark} \& {Murray}}{{Harper-Clark} \&
  {Murray}}{2009}]{Harper-Clark2009}
{Harper-Clark} E.,  {Murray} N.,  2009, \apj, 693, 1696

\bibitem[\protect\citeauthoryear{{Heyer}, {Krawczyk}, {Duval} \&
  {Jackson}}{{Heyer} et~al.}{2009}]{Heyer2009}
{Heyer} M.,  {Krawczyk} C.,  {Duval} J.,    {Jackson} J.~M.,  2009, \apj, 699,
  1092

\bibitem[\protect\citeauthoryear{{Hopkins}, {Quataert} \& {Murray}}{{Hopkins}
  et~al.}{2011}]{Hopkins2011}
{Hopkins} P.~F.,  {Quataert} E.,    {Murray} N.,  2011, \mnras, 417, 950

\bibitem[\protect\citeauthoryear{{Iliev}, {Ciardi}, {Alvarez}, {Maselli},
  {Ferrara}, {Gnedin}, {Mellema}, {Nakamoto}, {Norman}, {Razoumov},
  {Rijkhorst}, {Ritzerveld}, {Shapiro}, {Susa}, {Umemura} \& {Whalen}}{{Iliev}
  et~al.}{2006}]{Iliev2006}
{Iliev} I.~T.,  {Ciardi} B.,  {Alvarez} M.~A.,  {Maselli} A.,  {Ferrara} A.,
  {Gnedin} N.~Y.,  {Mellema} G.,  {Nakamoto} T.,  {Norman} M.~L.,  {Razoumov}
  A.~O.,  {Rijkhorst} E.-J.,  {Ritzerveld} J.,  {Shapiro} P.~R.,  {Susa} H.,
  {Umemura} M.,    {Whalen} D.~J.,  2006, \mnras, 371, 1057

\bibitem[\protect\citeauthoryear{{Iliev}, {Whalen}, {Mellema}, {Ahn}, {Baek},
  {Gnedin}, {Kravtsov}, {Norman}, {Raicevic}, {Reynolds}, {Sato}, {Shapiro},
  {Semelin}, {Smidt}, {Susa}, {Theuns} \& {Umemura}}{{Iliev}
  et~al.}{2009}]{Iliev2009}
{Iliev} I.~T.,  {Whalen} D.,  {Mellema} G.,  {Ahn} K.,  {Baek} S.,  {Gnedin}
  N.~Y.,  {Kravtsov} A.~V.,  {Norman} M.,  {Raicevic} M.,  {Reynolds} D.~R.,
  {Sato} D.,  {Shapiro} P.~R.,  {Semelin} B.,  {Smidt} J.,  {Susa} H.,
  {Theuns} T.,    {Umemura} M.,  2009, \mnras, 400, 1283

\bibitem[\protect\citeauthoryear{{Kim}, {Krumholz}, {Wise}, {Turk}, {Goldbaum}
  \& {Abel}}{{Kim} et~al.}{2013}]{Kim2013}
{Kim} J.-h.,  {Krumholz} M.~R.,  {Wise} J.~H.,  {Turk} M.~J.,  {Goldbaum}
  N.~J.,    {Abel} T.,  2013, \apj, 775, 109

\bibitem[\protect\citeauthoryear{{Krumholz} \& {Matzner}}{{Krumholz} \&
  {Matzner}}{2009}]{Krumholz2009}
{Krumholz} M.~R.,  {Matzner} C.~D.,  2009, \apj, 703, 1352

\bibitem[\protect\citeauthoryear{{Krumholz}, {Matzner} \& {McKee}}{{Krumholz}
  et~al.}{2006}]{Krumholz2006}
{Krumholz} M.~R.,  {Matzner} C.~D.,    {McKee} C.~F.,  2006, \apj, 653, 361

\bibitem[\protect\citeauthoryear{{Krumholz} \& {Thompson}}{{Krumholz} \&
  {Thompson}}{2012}]{Krumholz2012}
{Krumholz} M.~R.,  {Thompson} T.~A.,  2012, \apj, 760, 155

\bibitem[\protect\citeauthoryear{{Lagos}, {Lacey} \& {Baugh}}{{Lagos}
  et~al.}{2013}]{Lagos2013}
{Lagos} C.~d.~P.,  {Lacey} C.~G.,    {Baugh} C.~M.,  2013, \mnras, 436, 1787

\bibitem[\protect\citeauthoryear{{Levermore} \& {Pomraning}}{{Levermore} \&
  {Pomraning}}{1981}]{Levermore1981}
{Levermore} C.~D.,  {Pomraning} G.~C.,  1981, \apj, 248, 321

\bibitem[\protect\citeauthoryear{{Lopez}, {Krumholz}, {Bolatto}, {Prochaska},
  {Ramirez-Ruiz} \& {Castro}}{{Lopez} et~al.}{2013}]{Lopez2013}
{Lopez} L.~A.,  {Krumholz} M.~R.,  {Bolatto} A.~D.,  {Prochaska} J.~X.,
  {Ramirez-Ruiz} E.,    {Castro} D.,  2013, ArXiv e-prints 1309.5421

\bibitem[\protect\citeauthoryear{{Marinacci}, {Pakmor} \&
  {Springel}}{{Marinacci} et~al.}{2014}]{Marinacci2013}
{Marinacci} F.,  {Pakmor} R.,    {Springel} V.,  2014, \mnras, 437, 1750

\bibitem[\protect\citeauthoryear{{Mathews}}{{Mathews}}{1969}]{Mathews1969}
{Mathews} W.~G.,  1969, \apj, 157, 583

\bibitem[\protect\citeauthoryear{{Matzner}}{{Matzner}}{2002}]{Matzner2002}
{Matzner} C.~D.,  2002, \apj, 566, 302

\bibitem[\protect\citeauthoryear{{McKee}}{{McKee}}{1989}]{McKee1989}
{McKee} C.~F.,  1989, \apj, 345, 782

\bibitem[\protect\citeauthoryear{{McKee}, {van Buren} \& {Lazareff}}{{McKee}
  et~al.}{1984}]{McKee1984}
{McKee} C.~F.,  {van Buren} D.,    {Lazareff} B.,  1984, \apjl, 278, L115

\bibitem[\protect\citeauthoryear{{Mu{\~n}oz}, {Springel}, {Marcus},
  {Vogelsberger} \& {Hernquist}}{{Mu{\~n}oz} et~al.}{2013}]{Munoz2013}
{Mu{\~n}oz} D.~J.,  {Springel} V.,  {Marcus} R.,  {Vogelsberger} M.,
  {Hernquist} L.,  2013, \mnras, 428, 254

\bibitem[\protect\citeauthoryear{{Murray}, {Quataert} \& {Thompson}}{{Murray}
  et~al.}{2005}]{Murray2005}
{Murray} N.,  {Quataert} E.,    {Thompson} T.~A.,  2005, \apj, 618, 569

\bibitem[\protect\citeauthoryear{{Nelson}, {Vogelsberger}, {Genel}, {Sijacki},
  {Kere{\v s}}, {Springel} \& {Hernquist}}{{Nelson} et~al.}{2013}]{Nelson2013}
{Nelson} D.,  {Vogelsberger} M.,  {Genel} S.,  {Sijacki} D.,  {Kere{\v s}} D.,
  {Springel} V.,    {Hernquist} L.,  2013, \mnras, 429, 3353

\bibitem[\protect\citeauthoryear{{Oppenheimer} \& {Dav{\'e}}}{{Oppenheimer} \&
  {Dav{\'e}}}{2006}]{Oppenheimer2006}
{Oppenheimer} B.~D.,  {Dav{\'e}} R.,  2006, \mnras, 373, 1265

\bibitem[\protect\citeauthoryear{{Pakmor} \& {Springel}}{{Pakmor} \&
  {Springel}}{2013}]{Pakmor2013}
{Pakmor} R.,  {Springel} V.,  2013, \mnras, 432, 176

\bibitem[\protect\citeauthoryear{{Petkova} \& {Springel}}{{Petkova} \&
  {Springel}}{2009}]{Petkova2009}
{Petkova} M.,  {Springel} V.,  2009, \mnras, 396, 1383

\bibitem[\protect\citeauthoryear{{Renaud}, {Bournaud}, {Emsellem}, {Elmegreen},
  {Teyssier}, {Alves}, {Chapon}, {Combes}, {Dekel}, {Gabor}, {Hennebelle} \&
  {Kraljic}}{{Renaud} et~al.}{2013}]{Renaud2013}
{Renaud} F.,  {Bournaud} F.,  {Emsellem} E.,  {Elmegreen} B.,  {Teyssier} R.,
  {Alves} J.,  {Chapon} D.,  {Combes} F.,  {Dekel} A.,  {Gabor} J.,
  {Hennebelle} P.,    {Kraljic} K.,  2013, \mnras, 436, 1836

\bibitem[\protect\citeauthoryear{{Rogers} \& {Pittard}}{{Rogers} \&
  {Pittard}}{2013}]{Rogers2013}
{Rogers} H.,  {Pittard} J.~M.,  2013, \mnras, 431, 1337

\bibitem[\protect\citeauthoryear{{Sharma}, {Nath} \& {Shchekinov}}{{Sharma}
  et~al.}{2011}]{Sharma2011}
{Sharma} M.,  {Nath} B.~B.,    {Shchekinov} Y.,  2011, \apjl, 736, L27

\bibitem[\protect\citeauthoryear{{Shu}}{{Shu}}{1992}]{Shu1992v2}
{Shu} F.~H.,  1992, {The physics of astrophysics. Volume II: Gas dynamics.}

\bibitem[\protect\citeauthoryear{{Sijacki}, {Vogelsberger}, {Kere{\v s}},
  {Springel} \& {Hernquist}}{{Sijacki} et~al.}{2012}]{Sijacki2012}
{Sijacki} D.,  {Vogelsberger} M.,  {Kere{\v s}} D.,  {Springel} V.,
  {Hernquist} L.,  2012, \mnras, 424, 2999

\bibitem[\protect\citeauthoryear{{Spitzer}}{{Spitzer}}{1978}]{Spitzer1978}
{Spitzer} L.,  1978, {Physical processes in the interstellar medium}

\bibitem[\protect\citeauthoryear{{Springel}}{{Springel}}{2005}]{Springel2005b}
{Springel} V.,  2005, \mnras, 364, 1105

\bibitem[\protect\citeauthoryear{{Springel}}{{Springel}}{2010}]{Springel2010}
{Springel} V.,  2010, \mnras, 401, 791

\bibitem[\protect\citeauthoryear{{Stinson}, {Brook}, {Macci{\`o}}, {Wadsley},
  {Quinn} \& {Couchman}}{{Stinson} et~al.}{2013}]{Stinson2013}
{Stinson} G.~S.,  {Brook} C.,  {Macci{\`o}} A.~V.,  {Wadsley} J.,  {Quinn}
  T.~R.,    {Couchman} H.~M.~P.,  2013, \mnras, 428, 129

\bibitem[\protect\citeauthoryear{{Torrey}, {Vogelsberger}, {Genel}, {Sijacki},
  {Springel} \& {Hernquist}}{{Torrey} et~al.}{2014}]{Torrey2013}
{Torrey} P.,  {Vogelsberger} M.,  {Genel} S.,  {Sijacki} D.,  {Springel} V.,
  {Hernquist} L.,  2014, \mnras

\bibitem[\protect\citeauthoryear{{Vogelsberger}, {Genel}, {Sijacki}, {Torrey},
  {Springel} \& {Hernquist}}{{Vogelsberger} et~al.}{2013}]{Vogelsberger2013}
{Vogelsberger} M.,  {Genel} S.,  {Sijacki} D.,  {Torrey} P.,  {Springel} V.,
  {Hernquist} L.,  2013, \mnras, 436, 3031

\bibitem[\protect\citeauthoryear{{Vogelsberger}, {Sijacki}, {Kere{\v s}},
  {Springel} \& {Hernquist}}{{Vogelsberger} et~al.}{2012}]{Vogelsberger2012}
{Vogelsberger} M.,  {Sijacki} D.,  {Kere{\v s}} D.,  {Springel} V.,
  {Hernquist} L.,  2012, \mnras, 425, 3024

\bibitem[\protect\citeauthoryear{{Walch}, {Whitworth}, {Bisbas}, {W{\"u}nsch}
  \& {Hubber}}{{Walch} et~al.}{2012}]{Walch2012}
{Walch} S.~K.,  {Whitworth} A.~P.,  {Bisbas} T.,  {W{\"u}nsch} R.,    {Hubber}
  D.,  2012, \mnras, 427, 625

\bibitem[\protect\citeauthoryear{{Whitworth}}{{Whitworth}}{1979}]{Whitworth1979}
{Whitworth} A.,  1979, \mnras, 186, 59

\bibitem[\protect\citeauthoryear{{Wise}, {Abel}, {Turk}, {Norman} \&
  {Smith}}{{Wise} et~al.}{2012}]{Wise2012}
{Wise} J.~H.,  {Abel} T.,  {Turk} M.~J.,  {Norman} M.~L.,    {Smith} B.~D.,
  2012, \mnras, 427, 311

\bibitem[\protect\citeauthoryear{{Yorke}, {Tenorio-Tagle} \&
  {Bodenheimer}}{{Yorke} et~al.}{1983}]{Yorke1983}
{Yorke} H.~W.,  {Tenorio-Tagle} G.,    {Bodenheimer} P.,  1983, \aap, 127, 313

\end{thebibliography}


\appendix

\section{Standard Radiative Transfer Tests}

\begin{figure*}[H] 
\begin{center} 
\includegraphics[width=0.475\linewidth]{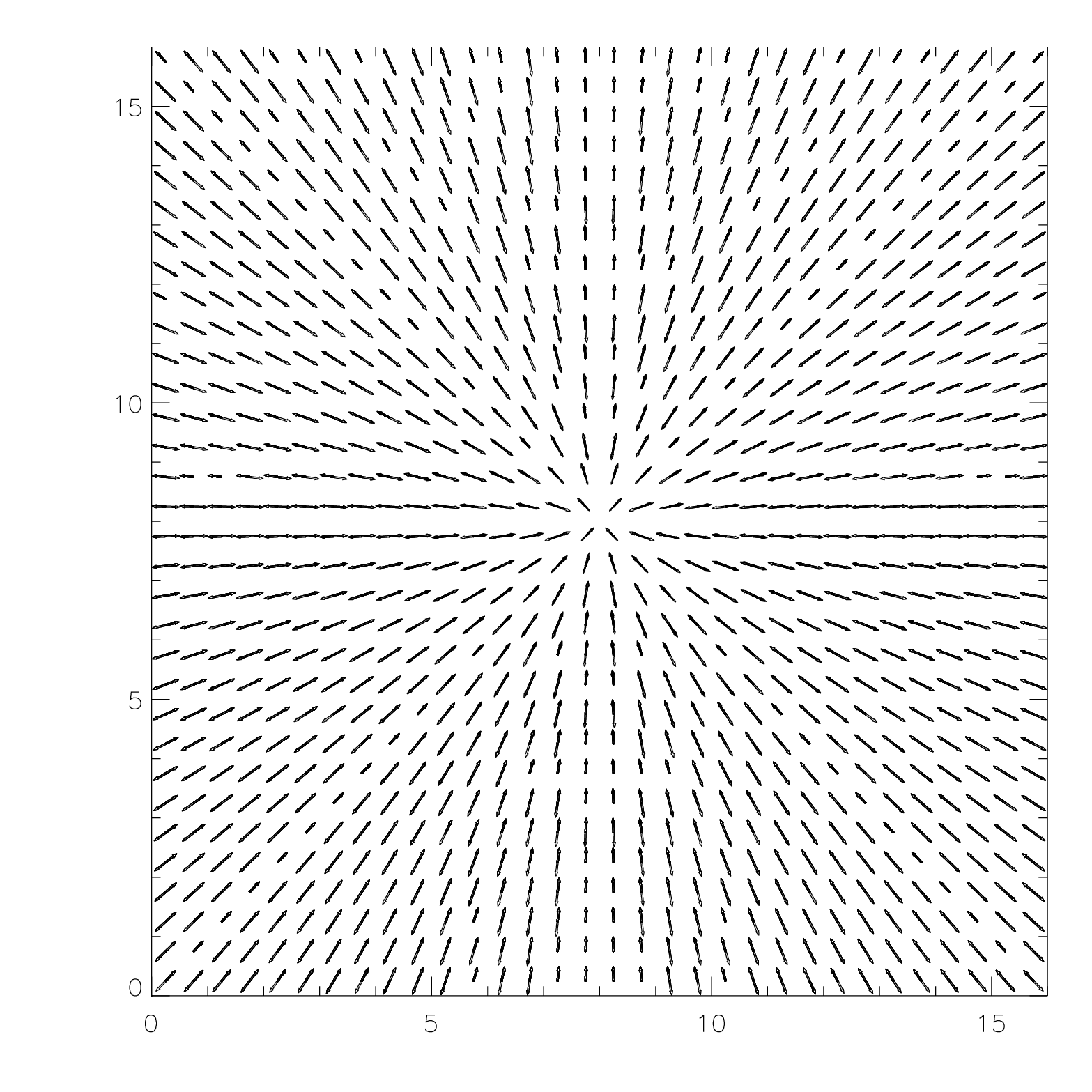} 
\includegraphics[width=0.475\linewidth]{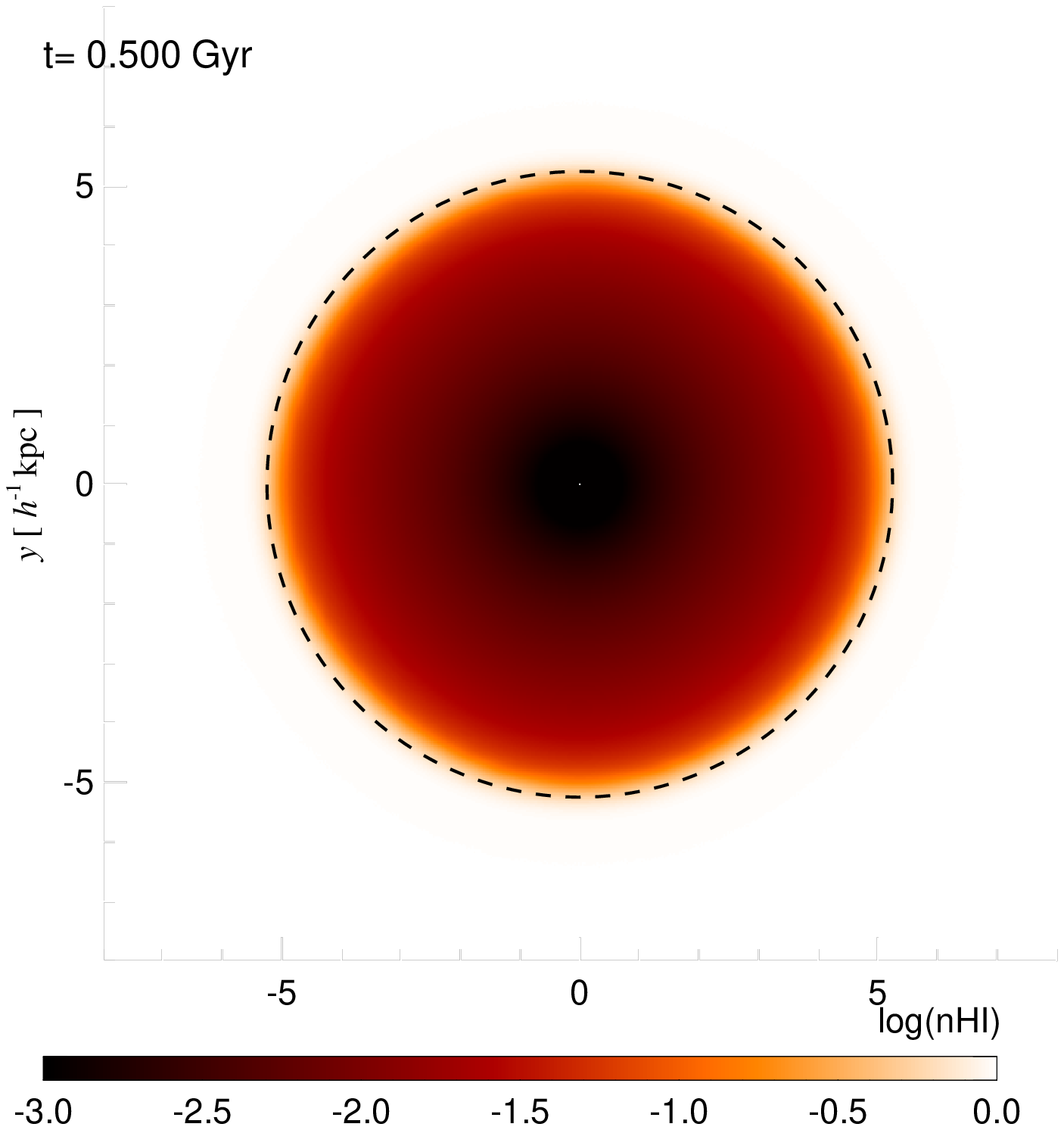} 
\caption{{\it Left:} Eigenvectors of the Eddington Tensor for a single
  central source in Test 1. {\it Right:} Map of neutral gas fraction
  for a central slab in a $64^3$ cells run after $t=500 \,\rm
  Myr$. The Str\"omgren radius is indicated with a black dashed line.}
\label{fig_rt:t1_img}
\end{center}
\end{figure*}

In this section we report the performance of the radiative transfer
module for the set of tests proposed in the radiative transfer code
comparison work by Iliev et al. (2006), augmented with the 
dynamical ``Test 5'' in Iliev et al. (2009). The initial set-ups are
the ones used previously in the {\sc gadget} version of the module
presented in Petkova \& Springel (2009). Refinement/de-refinement of
cells is not required in these tests as they deal mostly with static
and equal-density gas configurations, except for the last
test. Although the code has the capability to work with a mixture of
hydrogen and helium gas, in the interest of simplifying a comparison
with previous work, the gas is assumed to be composed of hydrogen only.

\subsection{Test 1: Pure hydrogen isothermal HII region expansion}

We study the ionization profiles in a constant density box with
$L_{\rm box} = 16 \, \rm kpc$ on a side. We use a central source
emitting $5 \times 10^{48}\,\rm photons \, s^{-1}$. The temperature is
fixed at $\rm T= 10^4 \, \rm K$, and the gas density is $n_{\rm
  H}=10^{-3} \, \rm cm^{-3}$. The radiation is mono-chromatic with
energy $E_0=13.6 \, \rm eV$. For the numbers above, the recombination
time is $t_{\rm rec}=125 \,{\rm Myr}$ and the Str\"omgren radius is
$r_s=5.38\,{\rm kpc}$.

\begin{figure} [H]
\begin{center} 
\includegraphics[width=84mm]{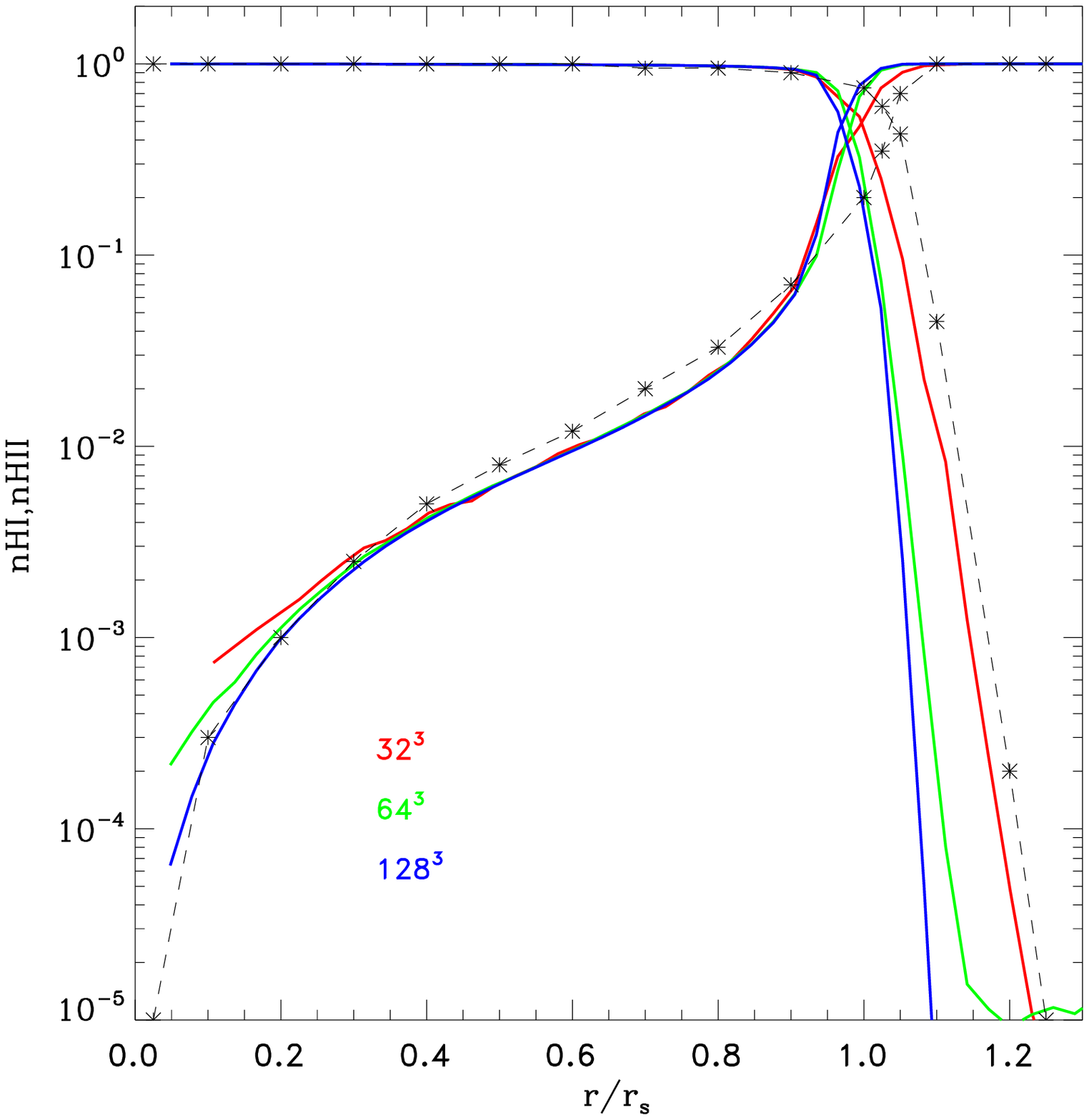} 
\caption{Neutral fraction profile for a single source in a box with
  $32^3$, $64^3$ and $128^3$ cells (Test 1). Dashed line/asterisks shows the
  theoretical solution taken from PS09. Overall, we find excellent
  convergence of results with the number of cells.}
\label{fig_rt:t1_ngrid}
\end{center}
\end{figure}

We show a projection of the (dominant) eigenvectors of the Eddington
tensor in a thin slab of the box (right panel of
Fig.~\ref{fig_rt:t1_img}). The tensor determines the direction of the
radiation transport, which is radially away from the source for our
specific case. The right panel in Fig.~\ref{fig_rt:t1_img} shows an
intensity map of the neutral gas fraction after $500\,{\rm Myr}$ of
evolution. The black dashed line indicates, as expected, that the
Str\"omgren radius coincides with the radius beyond which the gas
remains neutral.  We use $64^3$ cells for this experiment and test the
numerical convergence with the number of cells below.

Fig.~\ref{fig_rt:t1_ngrid} shows that the neutral/ionized gas profiles
are in good agreement with the theoretical predictions (dashed black
line connecting asterisks). We find that the flux-limiter determines
to a certain extent the shape of these profiles and use this fact to
calibrate the flux-limiter formula in our code.  We settled for the
\citet{Levermore1981} flux-limiter (see Eq.~\ref{eq:fluxlimiter}),
which in our case matches the analytical solutions better than the one
used previously in PS09. Different colours in
Fig.~\ref{fig_rt:t1_ngrid} show that the results converge well with
the number of cells ($32^3$ in red, $64^3$ in green and $128^3$ in
blue).

\begin{figure} 
\begin{center} 
\includegraphics[width=84mm]{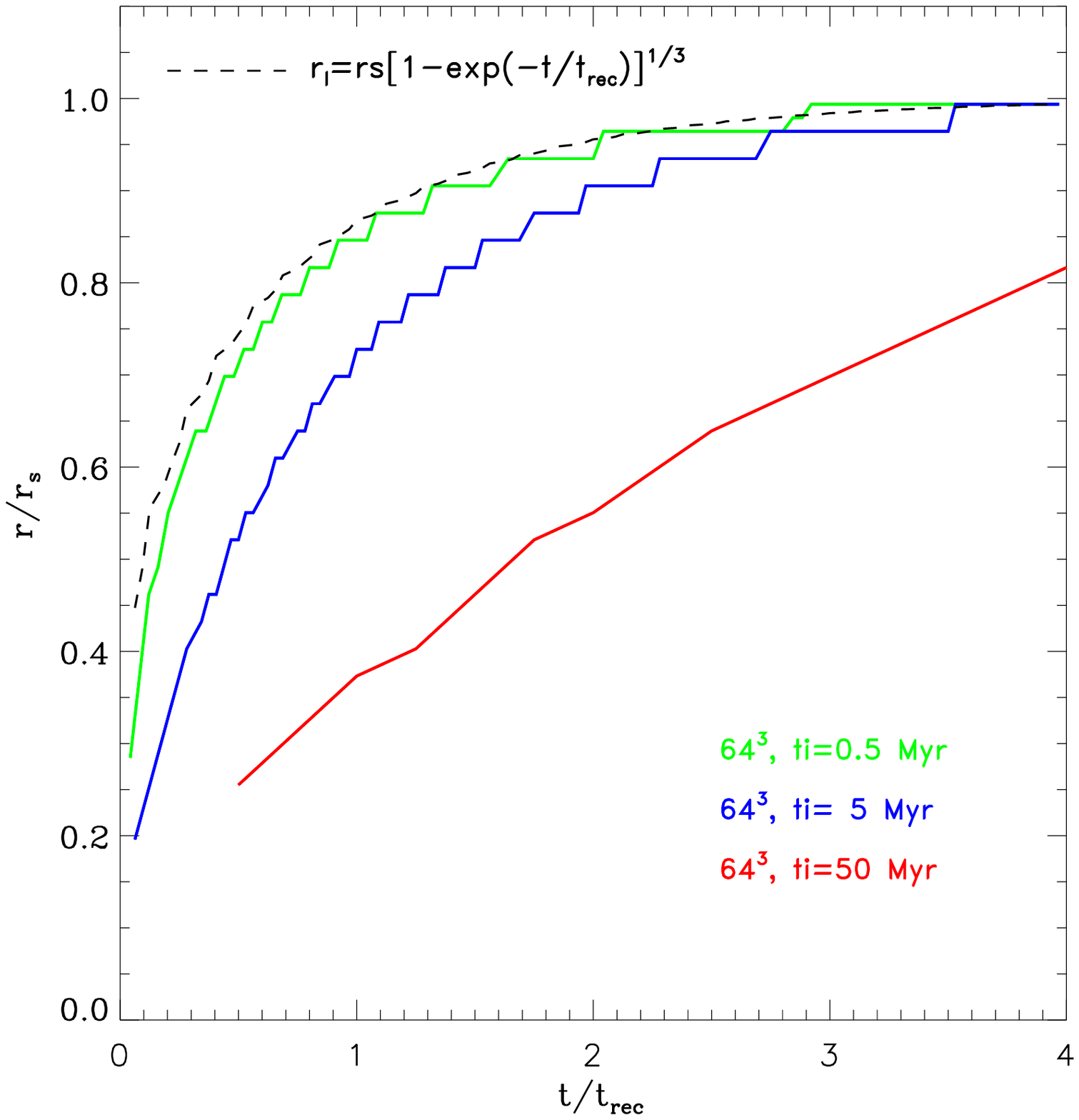} 
\caption{Effect of time-step size on the I-front propagation. We use a
  $64^3$ grid. A time-step shorter than $0.5\,\rm Myr$ nicely
  reproduces the expected theoretical results (dashed black curve).}
\label{fig_rt:t1_dt}
\end{center}
\end{figure}

We explore the effects of different time-stepping in
Fig.~\ref{fig_rt:t1_dt}. We show the time evolution of the ionization
radius for this test with fixed individual (global) time-steps $\Delta t =50$,
$5$ and $0.5 \, \rm Myr$. The black dashed line shows the result from
the analytic formula in Eq.~(\ref{eq:ri_time}). The agreement improves
for runs with small time-steps, which are able to better resolve the
early evolution of the front. For instance, for $\Delta t=5 \, \rm
Myr$, the simulated and analytical results agree well only after
$t\sim 3 \, t_{\rm rec}$; instead, a simulation with 10 times smaller
time-steps describes the whole time evolution of the ionization front
accurately.

\begin{figure*} 
\begin{center} 
\includegraphics[width=0.475\linewidth]{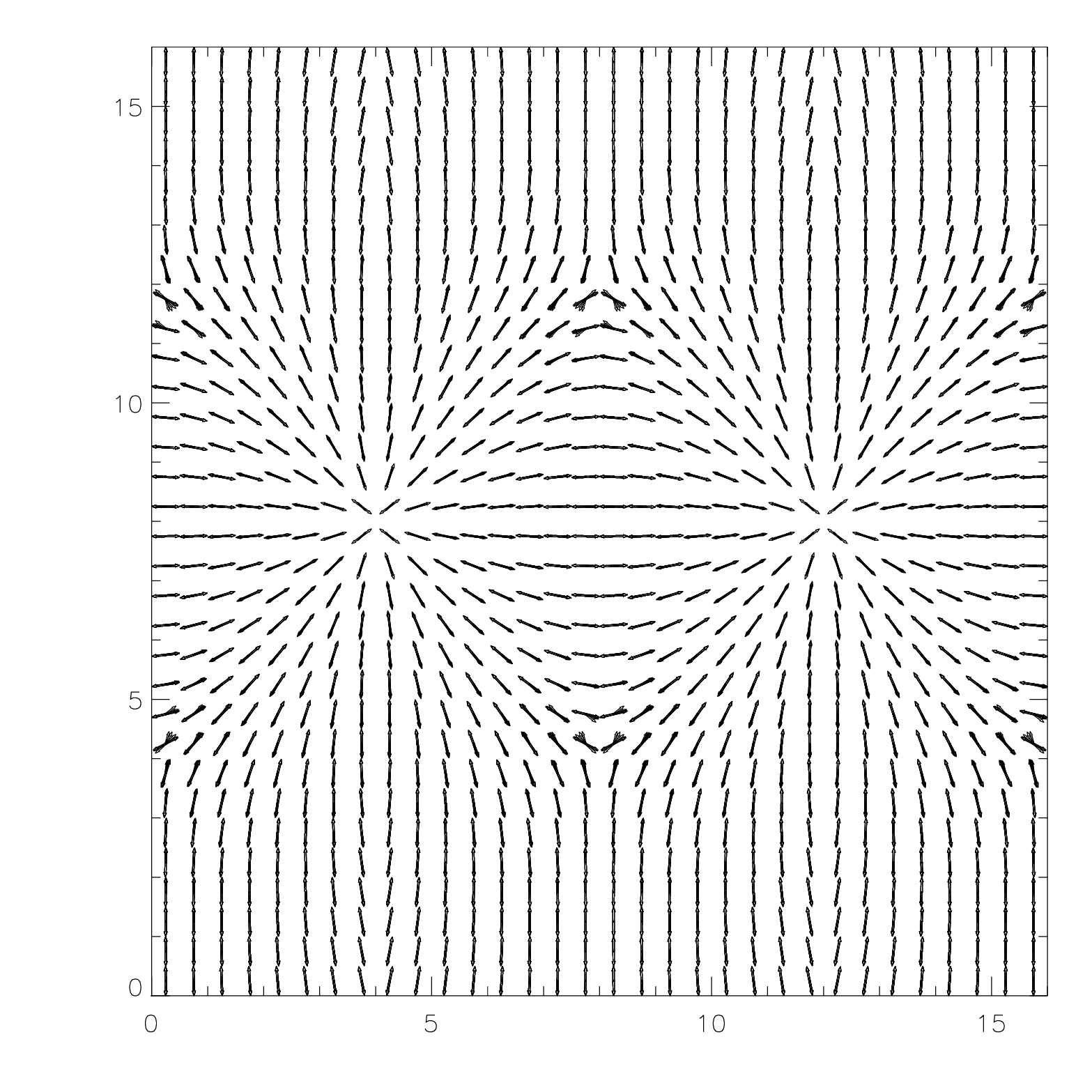} 
\includegraphics[width=0.475\linewidth]{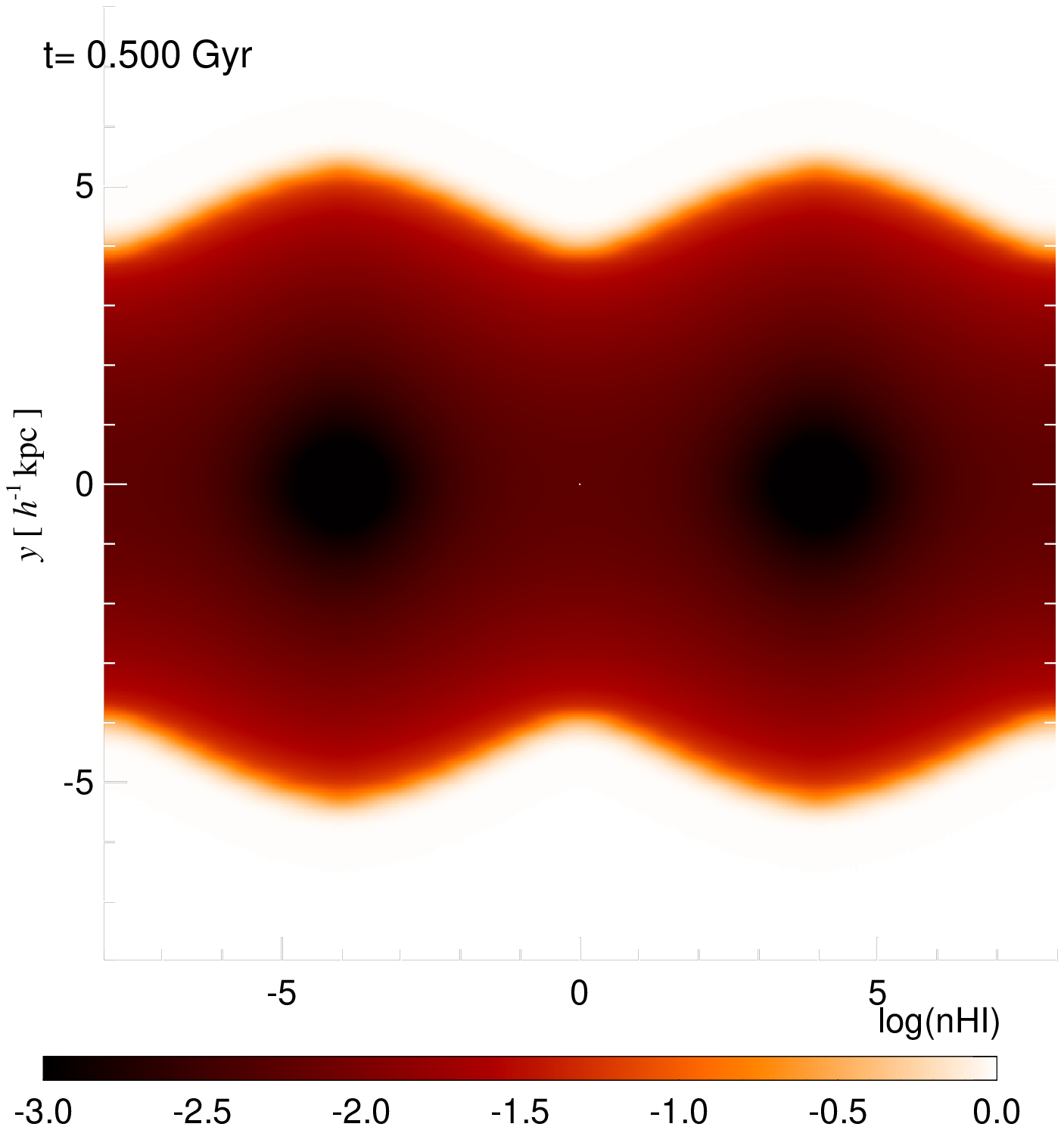} 
\caption{Same as Fig.~\ref{fig_rt:t1_img} but for two nearby
  sources. We use $64^3$ cells.}
\label{fig_rt:t1_img2}
\end{center}
\end{figure*}

Finally, we also explored a double-source set up as suggested by
\citet{Gnedin2001} and PS09. This test has the same parameters for gas
and source as before, but we employ two sources (instead of one)
separated by $8 \, \rm kpc$.  Their Str\"omgren spheres do overlap and
we end up with a shared, elongated ionized bubble after $500 \, \rm
Myr$, as shown in the right panel of Fig.~\ref{fig_rt:t1_img2}.  Our
results agree well with those in PS09.

\subsection{Test 2: HII region expansion: the temperature field}

This test consists of a static uniform density field within a box of
side-length $16$ kpc, in which we place a central source emitting $5
\times 10^{48}\, \rm photons\,s^{-1}$ with a black-body spectrum of
temperature $T=10^5 \, \rm K$.  Initially, the gas is set to
density and temperature equal to $n_{\rm H}=10^{-3} \, \rm cm^{-3}$
and $T= 10^2 \, \rm K$, respectively, with the latter being allowed to
change according to the heating and cooling mechanisms described in
Sec~\ref{sec:rt}.

\begin{figure} 
\begin{center} 
\includegraphics[width=84mm]{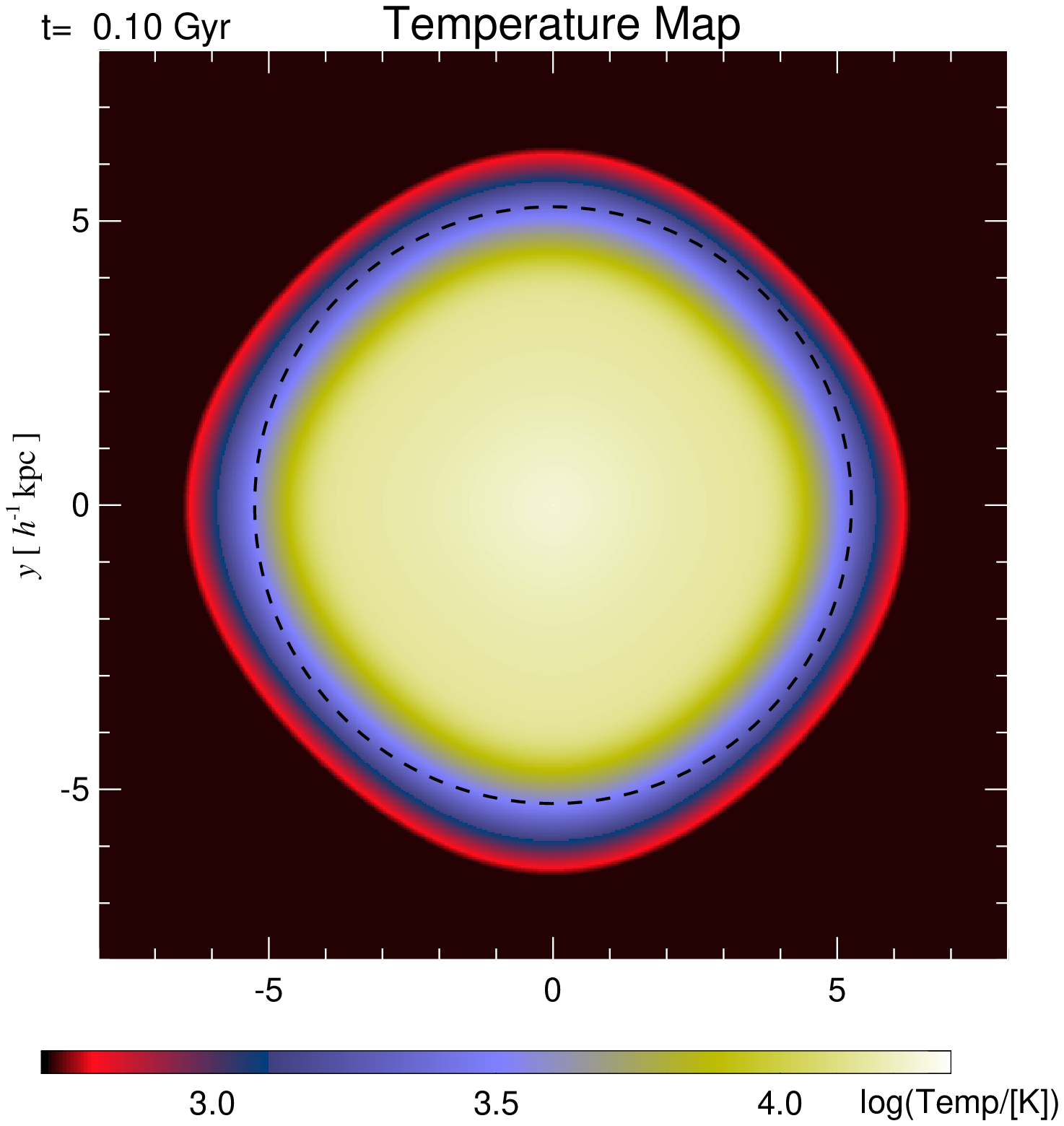} 
\caption{Temperature map corresponding to Test 2 after $t=100\,\rm
  Myr$.  For guidance, we show the Str\"omgren radius from Test 1 with
  a dashed black line.
\label{fig_rt:t2_tmap}}
\end{center}
\end{figure}

\begin{figure} 
\begin{center}
\includegraphics[width=84mm]{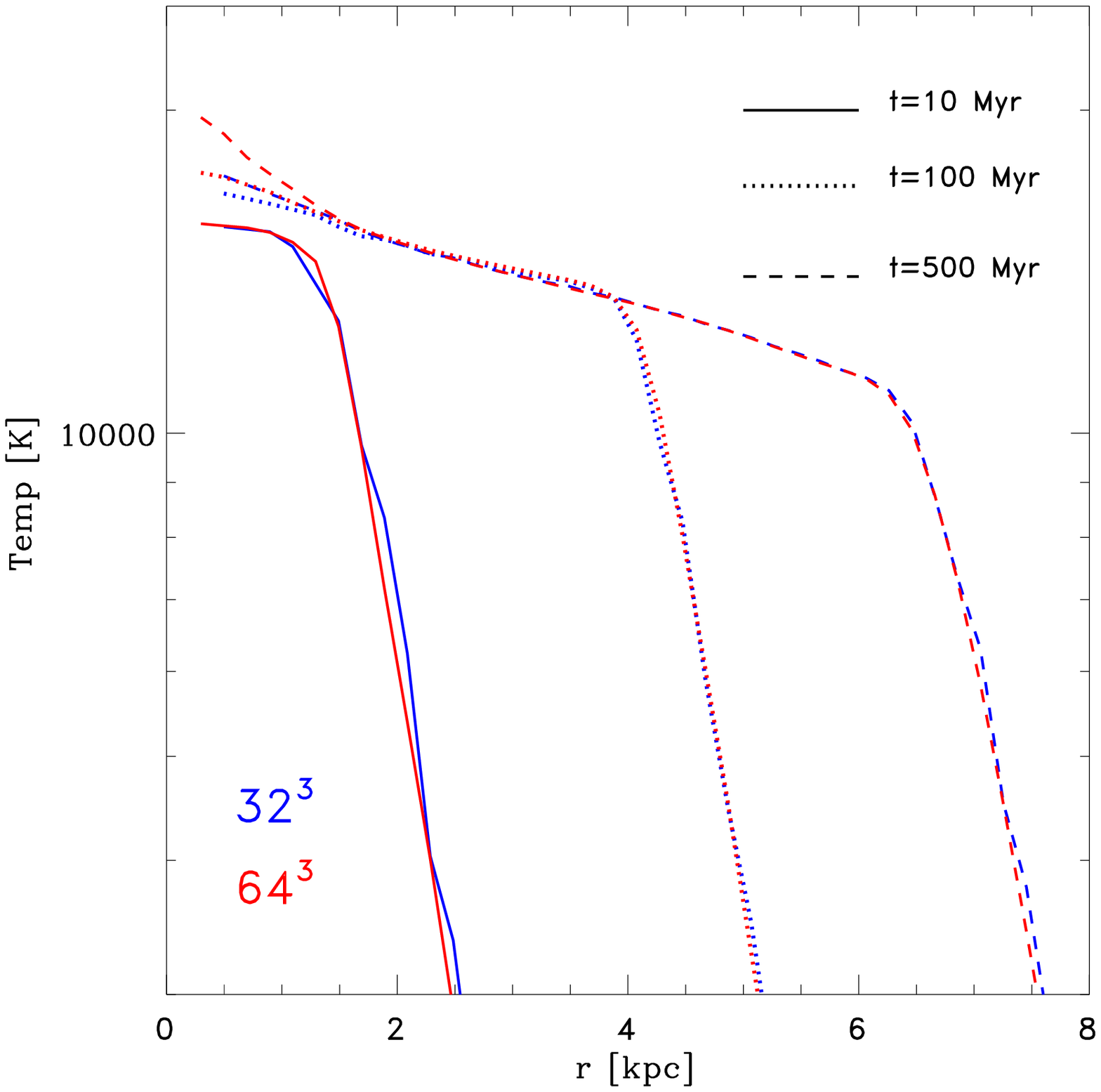} 
\caption{Spherically averaged temperature profiles in the HII region
  from Test 2 at three different times. We obtain good convergence of
  the results when varying the number of cells ($32^3$ in blue, $64^3$
  in red).}
\label{fig_rt:t2_t_r}
\end{center}
\end{figure}

Fig.~\ref{fig_rt:t2_tmap} shows a temperature map in a central slice
through the simulated box at $t=0.1 \, \rm Myr$. Unlike in the
previous test, a set-up with varying temperature does not have an
analytical solution for the size of the Str\"omgren sphere. For
reference we also show the Str\"omgren radius corresponding to Test 1,
that has the same conditions but fixed $T=10^4 \, \rm K$ (dashed black
line). Photoionization proceeds slightly beyond this estimate,
probably due to the presence of high-energy photons that are able to
penetrate deeper into the neutral gas.

\begin{figure} 
\begin{center} 
\includegraphics[width=84mm]{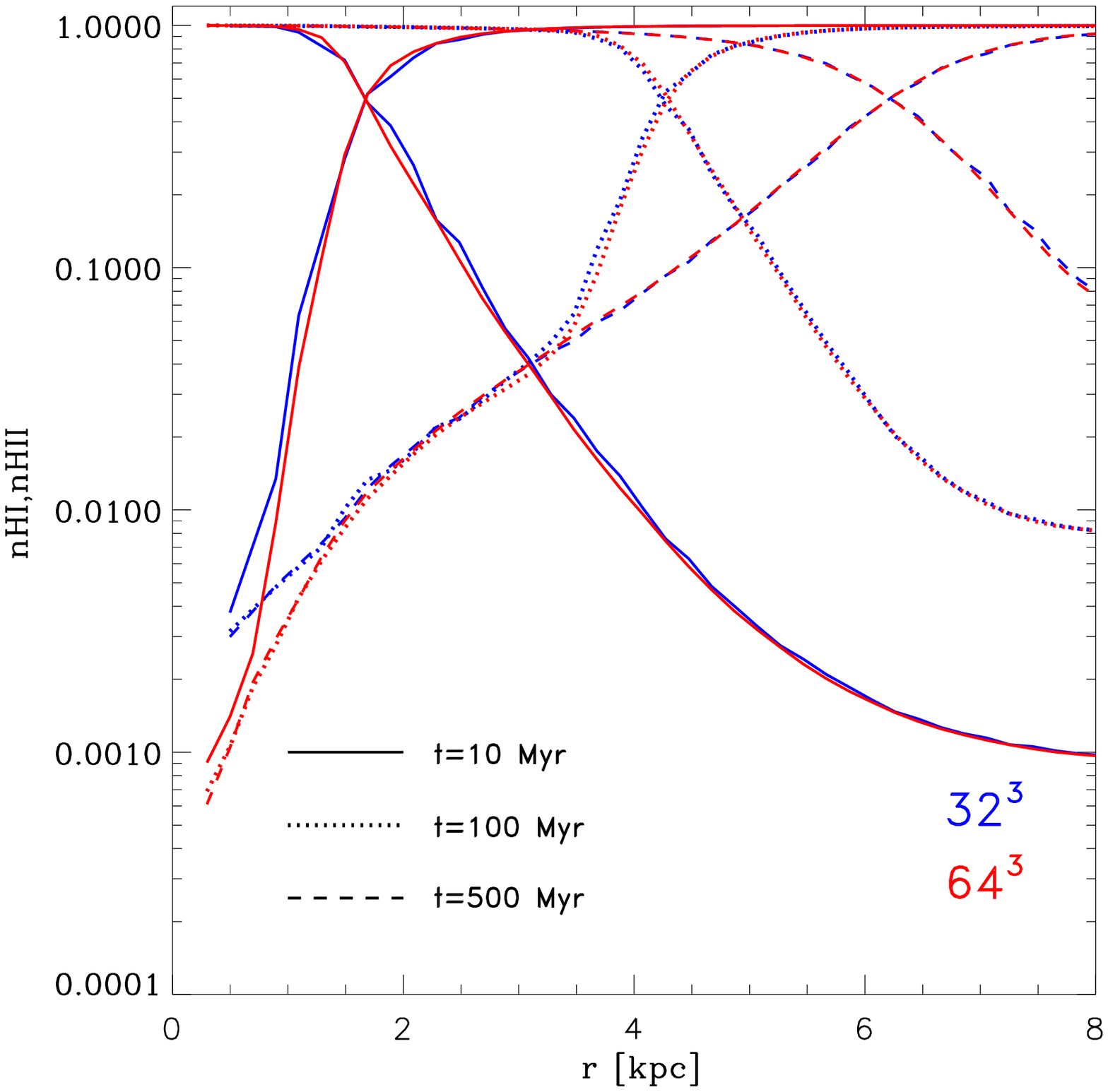} 
\caption{Same as Fig.~\ref{fig_rt:t2_t_r}, but for the neutral/ionized
gas profile.}
\label{fig_rt:t2_nH_r}
\end{center}
\end{figure}

Fig.~\ref{fig_rt:t2_t_r} shows in more detail the temperature profiles
at three different times $t=10$, $100$ and $500 \, \rm Myr$, and for
two different resolutions corresponding to $32^3$ and $64^3$
cells. The results show excellent numerical convergence and their
behaviour with time and radius agrees well with those reported in PS09
and Iliev et al.~(2006). The same is true for the neutral/ionized gas
profiles shown in Fig.~\ref{fig_rt:t2_nH_r}. Here the action of more
energetic photons, which smooth out the transition between ionized and
neutral region compared to a mono-chromatic source like in Test~1, can
be clearly appreciated.

\begin{figure*} 
\begin{center} 
\includegraphics[width=0.33\linewidth]{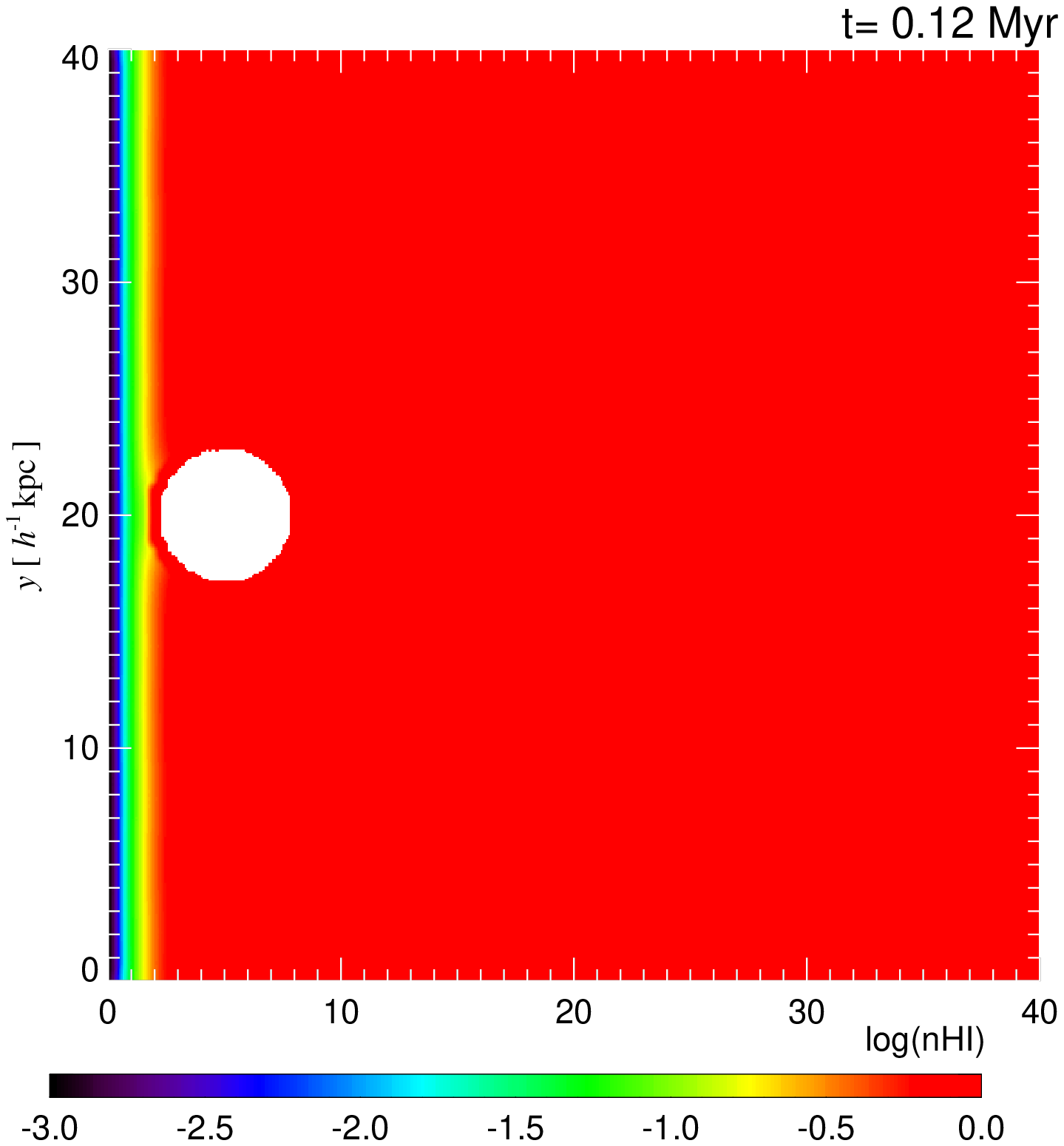} 
\includegraphics[width=0.33\linewidth]{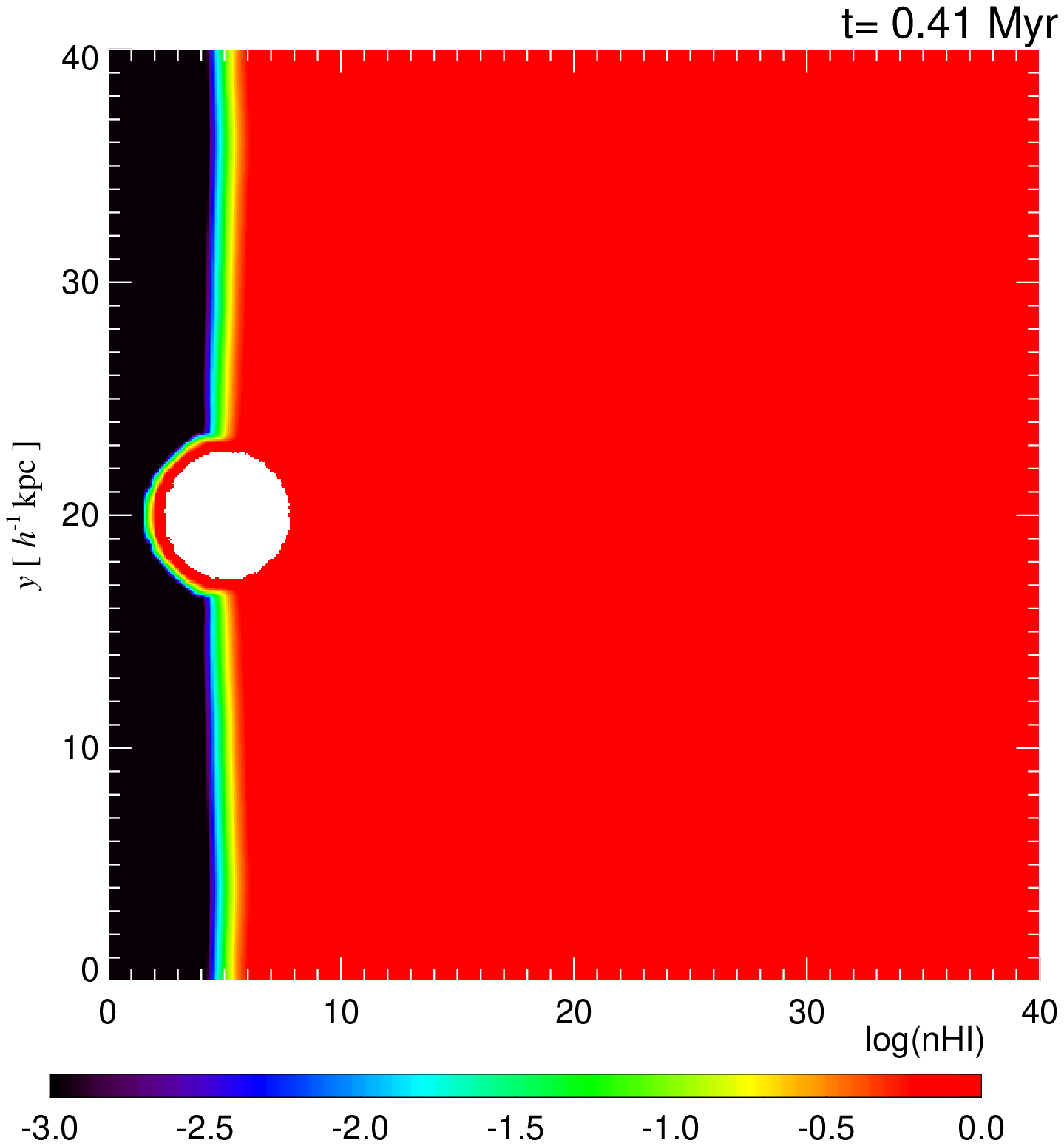} 
\includegraphics[width=0.33\linewidth]{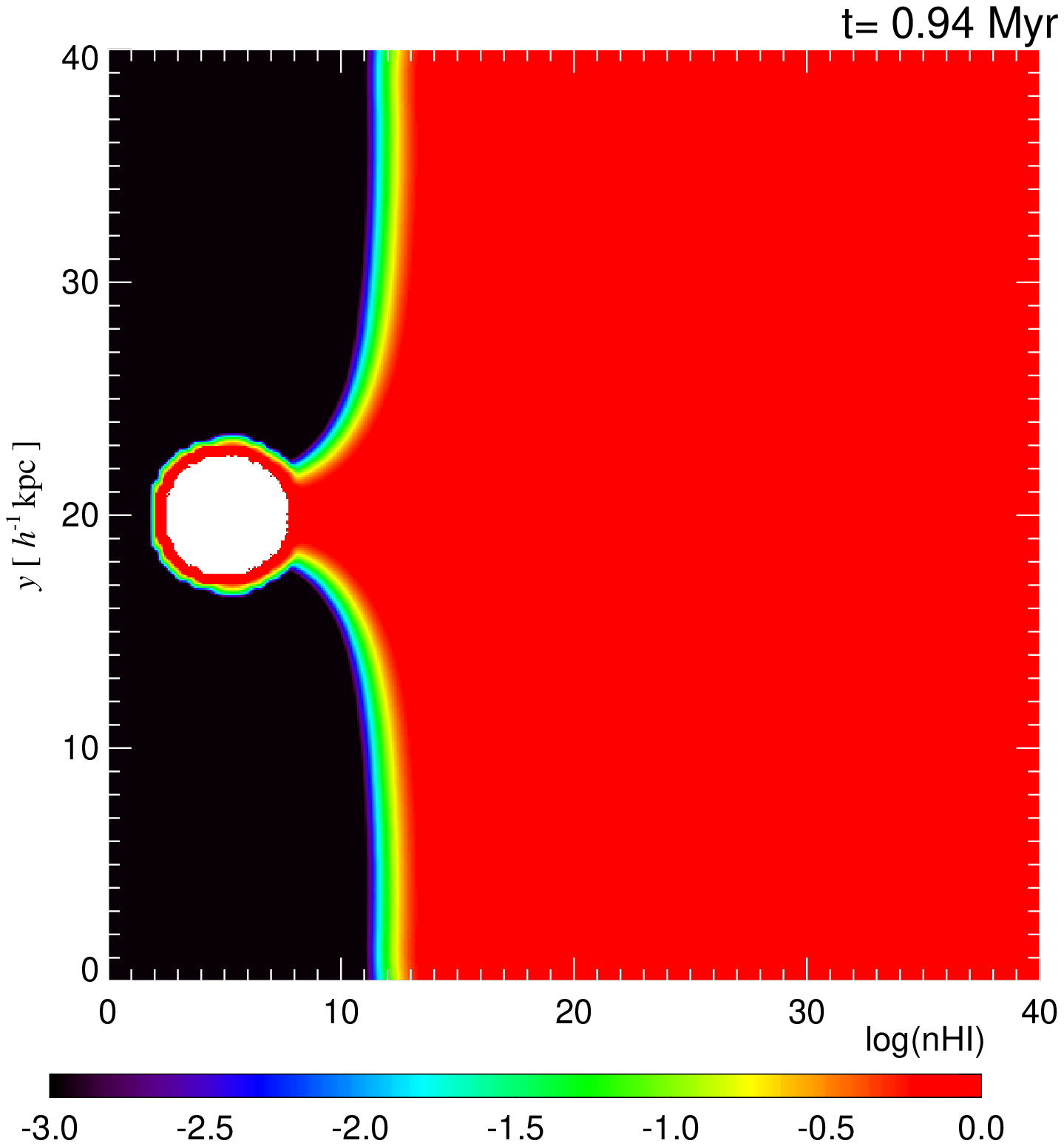} 
\caption{Shadow test: Ionized fraction in a slide through the middle
  of the box at three different times (Test 3). The evolution is as
  expected, but notice that the OTVET approximation fails to cast a
  sharp shadow after the I-front has advanced beyond the center of the
  clump. This extra diffusion is expected and a well known issue of
  this method.}
\label{fig_rt:t3}
\end{center}
\end{figure*}

\subsection{Test 3: I-front trapping in a dense clump and the
formation of a shadow}

This set-up follows exactly the test presented in PS09, which in
turn is inspired by Test 3 in Iliev et
al.~(2006). It consists of a box with length 40 kpc on a side filled
with gas at density $n_{\rm H}=10^{-3} \, \rm cm^{-3}$ that contains a
cylinder $10^5$ times denser. Radiation comes from a plane of (512)
stars located in the $x$-$y$ plane, each emitting $1.2 \times
10^{50}\, \rm photons\, s^{-1}$. The center of the cylinder is located
at $(x,y) = (5\, {\rm kpc},\,20 \, {\rm kpc})$ and is aligned with the
$z$-axis.

We show the time progression of the test in Fig.~\ref{fig_rt:t3}. The
ionization front should move from left to right with time in this
projection. At the beginning of the simulation, the clump (seen here
in white) successfully halts the progress of ionization into the high
density gas. However, after $\sim 0.5 \, \rm Myr$ the ionizing
radiation is able to penetrate into the region behind the clump,
failing to create the expected sharp ``shadow''.  This feature is a
well known problem of moment-based radiative transfer schemes
\citep{Gnedin2001, Aubert2008, Petkova2009} and can be explained by the
residual contributions of cells not directly aligned with the
source. The Eddington tensors of such cells have more than one
non-zero component (in this example, radiation should only propagate
in the $x$-direction), seeding the diffusion of the radiation into the
supposedly shadowed region.

\begin{figure*} [H]
\begin{center} 
\includegraphics[width=0.475\linewidth]{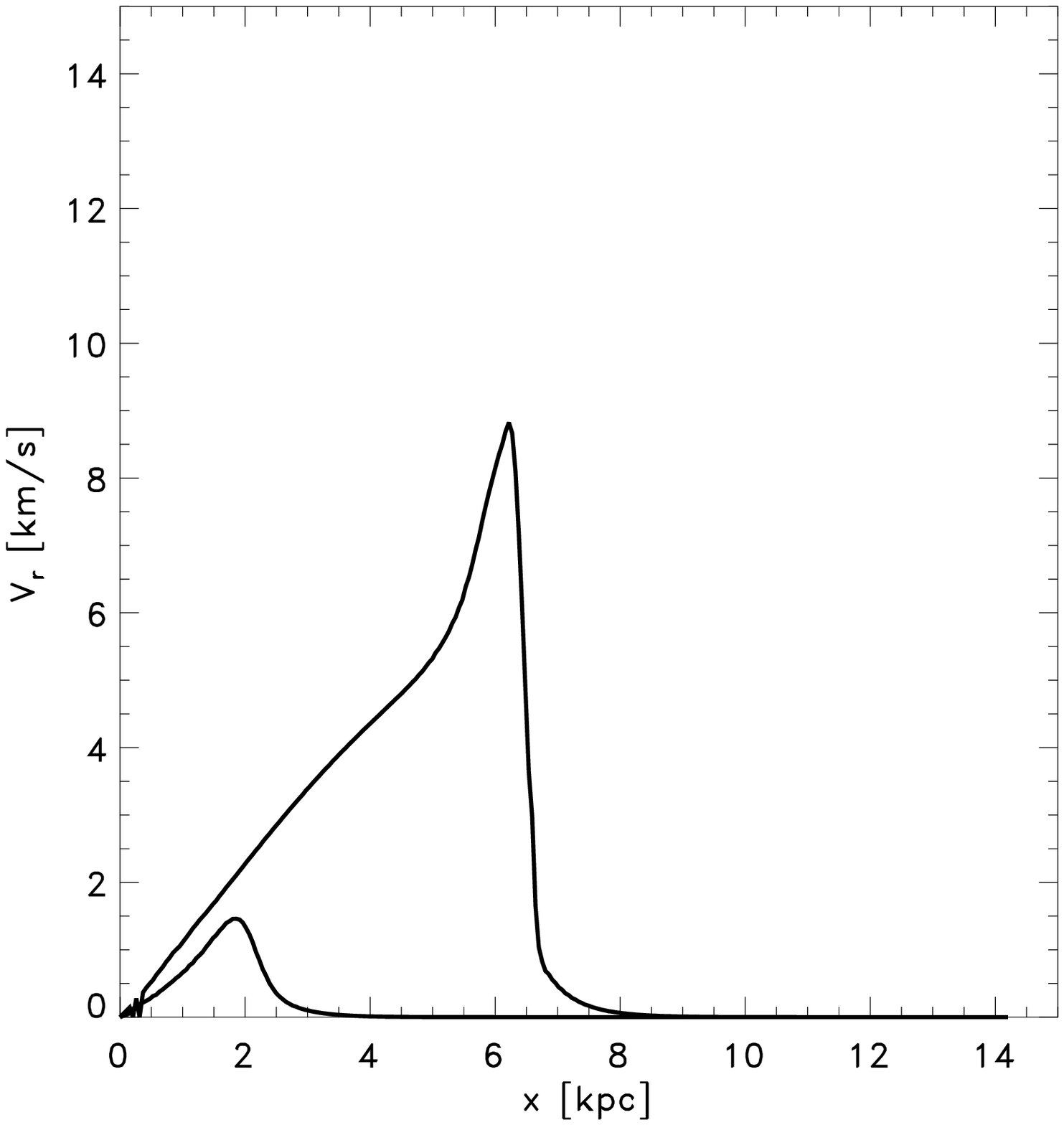} 
\includegraphics[width=0.475\linewidth]{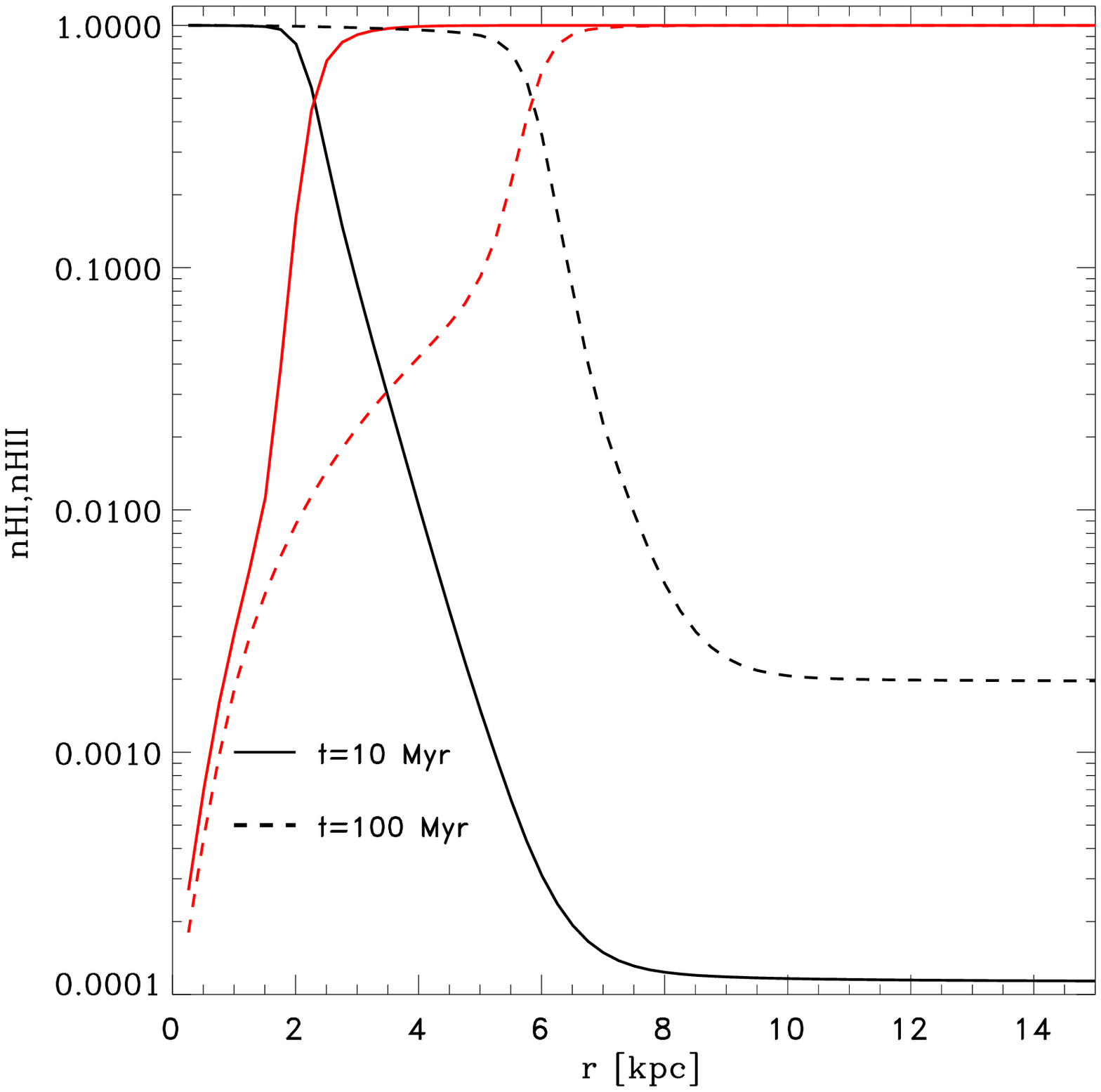} 
\caption{Radial velocity (left) and ionization profiles (right) for an
  HII expansion test equivalent to Test 5 in Iliev et al. (2009). We
  show times $t=10$ and $100 \, \rm Myr$.}
\label{fig_rt:t5}
\end{center}
\end{figure*}

Notice that we do not expect our results to be strongly affected by
the sub-optimal performance of the code in shadowing tests, as our
science applications of the radiative transfer module deal with
spherically symmetric configurations in a homogeneous medium.

\subsection{Test 4:  classical HII region expansion}

This test corresponds to Test 5 suggested in \citet{Iliev2009}.
Unlike the previous experiments, it studies the {\it dynamical}
response of the gas to ionization; thus this test is of particular
interest for our study. Following Illiev et al., we set up a constant
density box with $n_{\rm H}=10^{-3} \, \rm cm^{-3}$ and temperature
$T=10^2 \, \rm K$ where we place a central source emitting at a rate
$5 \times 10^{48} \, \rm s^{-1}$ photons per second with a $T_{\rm
  BB}=10^5 \, \rm K$ black-body spectrum.  Iliev et al. originally
used a box size of 15 kpc with 128 resolution elements, placing the
source in a corner and treating the boundaries of the box as periodic
or non-periodic according to their location.  For simplicity, we
prefer to deal with a central source and treat all boundaries as
non-periodic. We therefore use double the size of the box as well as
twice the number of cells ($L=30\,{\rm kpc}$ and $256^3$,
respectively) to match the original numerical resolution.

We find a very good agreement between our results and those in the
code comparison paper. For brevity, we show only the velocity and
ionization profiles in Fig.~\ref{fig_rt:t5} for $t=10$ and $100\,{\rm
  Myr}$, but we have checked that the agreement extends also to the
other properties such as temperature, density and pressure profiles.

Summarizing, our code, as all moment-based methods, tends to be too
diffusive, despite the efforts to capture the anisotropic propagation
of photons. As discussed in PS09, this might affect the geometry of
ionized regions in cases where the gas presents a large degree of
inhomogeneities. However, apart from this defect, the successful
performance of the code in several standard test problems suggests
that general properties of the ionized bubbles, such as volume, size,
temperature, pressure structure and induced gas dynamics should be
properly captured by our scheme.

\section{Convergence with number of cells}

\begin{figure} [H]
\begin{center} 
\includegraphics[width=84mm]{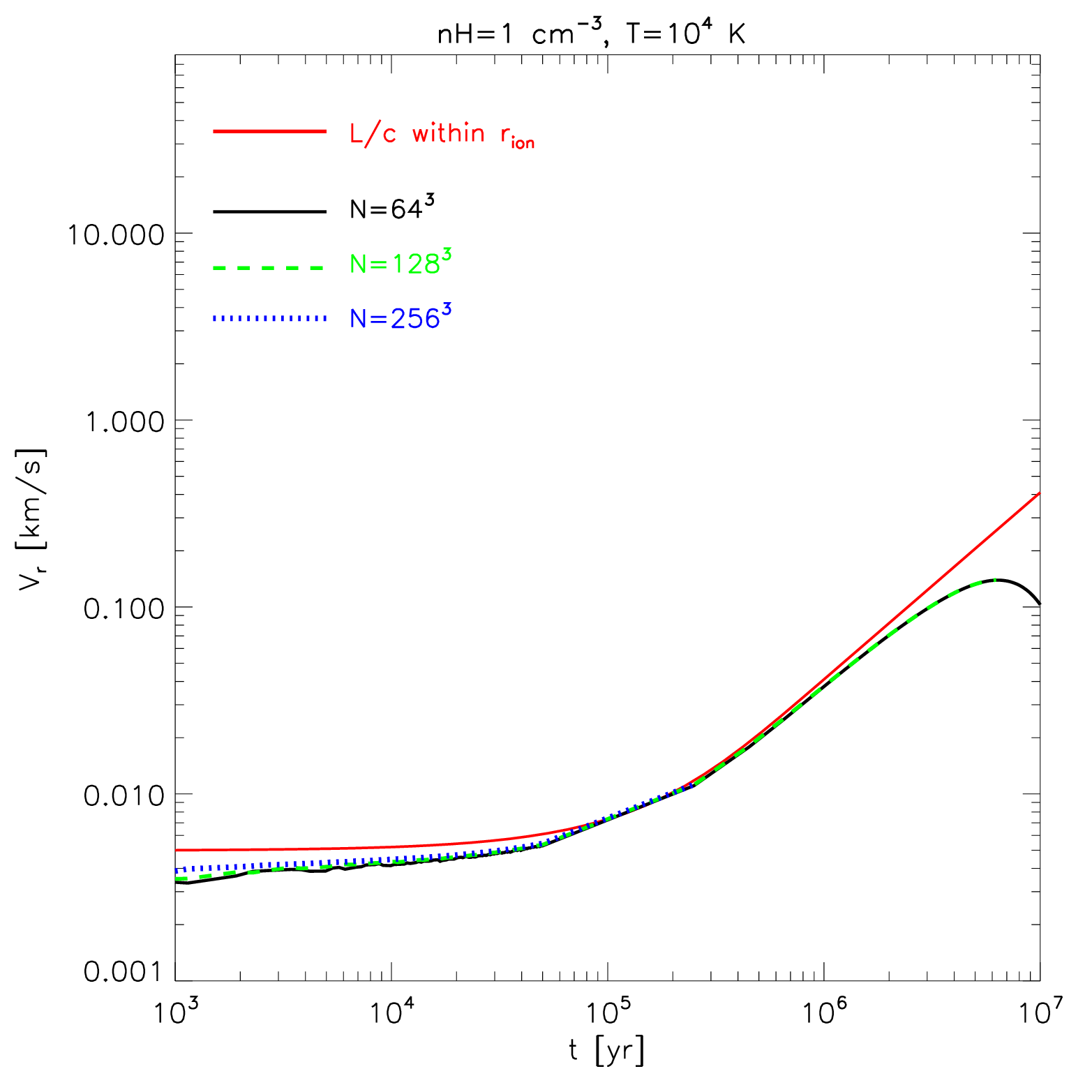} 
\caption{Convergence of the gas velocity with the number of cells,
  measured in a constant density box with $n_{\rm H}=1 \, \rm
  cm^{-3}$, $T=10^4 \, \rm K$ and luminosity $L = 10^6 \, \rm
  L_\odot$. The red solid curve indicates the analytical estimate from
  Eq.~(\ref{eq:lc}).
\label{fig:resolution}}
\end{center}
\end{figure}

Fig.~\ref{fig:resolution} shows the numerical convergence of our
radiation pressure results with the number of cells. We show the
radial velocity measured in the gas for a constant density box with
density $n_{\rm H}=1 \, \rm cm^{-3}$ and temperature $T=10^4 \, \rm
K$. In general, we use $64^3$ cells in our constant density
experiments of Sec.~\ref{sec:constant}.  Here, we compare $64^3$
against the same set-up using $128^3$ and $256^3$ cells. We find
excellent agreement between the curves, indicating that our results
are not affected by resolution effects. Notice that, as discussed in
the main text, Eq.~(\ref{eq:lc}) tends to overestimate the gas
velocity at early times. This effect is small, but most importantly,
it is independent of our numerical resolution.

\end{document}